\documentclass[printer]{tJOT2e}

\usepackage{epsfig}
\usepackage{dsfont}
\usepackage{amssymb}
\usepackage{amsmath}

\newcommand{\be}{\begin{equation}}
\newcommand{\en}{\end{equation}}
\newcommand{\bea}{\begin{eqnarray}}
\newcommand{\ena}{\end{eqnarray}}

\usepackage{color}
\newcommand{\rem}[1]{}

\begin{document}

\doi{10.1080/14685240xxxxxxxxxxxxx} \issn{1468-5248} \issnp{}
\jvol{00} \jnum{00} \jyear{2006} \jmonth{January}

\markboth{Mixing in manipulated turbulence}{Mixing in manipulated
turbulence}

\title{Mixing in manipulated turbulence}

\author{Arkadiusz K. Kuczaj$^{(1)}$ \thanks{To whom correspondence should
be addressed (a.k.kuczaj@utwente.nl)}, Bernard J.
Geurts$^{(1,2)}$ \\
\begin{enumerate}
\item[(1)] Multiscale Modeling and Simulation, NACM, J.M. Burgers Center, Faculty EEMCS, University of Twente, P.O. Box 217, 7500 AE Enschede, The Netherlands
\item[(2)] Anisotropic Turbulence, Fluid Dynamics Laboratory, Department of Applied Physics, Eindhoven University of Technology, P.O. Box 513, 5600 MB Eindhoven, The Netherlands
\end{enumerate}}

\received{\today}

\maketitle

\begin{abstract}
A~new computational framework for the simulation of turbulent flow
through complex objects and along irregular boundaries is presented.
This is motivated by the~application of metal foams in compact
heat-transfer devices, or as catalyst substrates in
process-engineering. The~flow-consequences of such complicated
objects are incorporated by adding explicit multiscale forcing to
the~Navier--Stokes equations. The~forcing represents
the~simultaneous agitation of a~wide spectrum of length-scales when
flow passes through the~complex object. Two types of forcing
procedures are investigated; with reference to the collection of
forced modes these procedures are classified as `constant-energy' or
`constant-energy-input-rate'. It is found that a~considerable
modulation of the traditional energy cascading can be introduced
with a specific forcing strategy. In spectral space, forcing yields
{\it{strongly localized}} deviations from the common Kolmogorov
scaling law, directly associated with the explicitly forced
scales. In~addition, the accumulated effect of forcing induces a~significant {\it{non-local}}
alteration of the~kinetic energy including the spectrum for the
large scales. Consequently, a manipulation of turbulent flow can be
achieved over an extended range, well beyond the directly forced
scales. Compared to flow forced in the large scales only, the~energy
in broad-band forced turbulence is found to be transferred more
effectively to smaller scales. The~turbulent mixing of a~passive
scalar field is also investigated, in order to quantify the
physical-space modifications of transport processes in multiscale
forced turbulence. The~surface-area and wrinkling of level-sets of
the scalar field are monitored as measures of the~influence of
explicit forcing on the local and global mixing efficiency. At small
Schmidt numbers, the~values of surface-area are mainly governed by
the~large scale sweeping-effect of the~flow while the~wrinkling is
influenced mainly by the~agitation of the~smaller
scales.\\

\noindent This paper is associated with the focus-issue
{\it{Multi-scale Interactions in Turbulent Flows.}}

\end{abstract}

\section{Introduction}
\label{intro}

\begin{figure}[hbt]
\centering{
\includegraphics[width=0.45\textwidth]{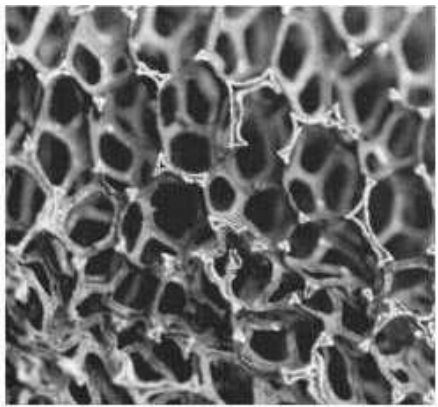}
} \caption{A porous nickel foam contains various geometrical
complexities on different length-scales \cite{li:study:2005}.}
\label{foam}
\end{figure}

Various multiscale phenomena in turbulent flows arise from the
passage of fluid through and along geometrically complex objects
placed inside the flow-domain. The corresponding perturbations of
the~flow arise simultaneously on a~range of length-scales and find
their origin in the complexity of the boundaries of these objects.
A~motivating example is the flow through a~porous region such as a
metal foam depicted in Fig.~\ref{foam}. Many more examples can
readily be mentioned, arising in different technological
applications or in numerous natural flows, including flow over
forest
canopies~\cite{finnigan:turbulence:2000,banhart:cellular:2001,boomsma:metal:2002}.

The~purpose of this paper is to investigate the~computational
modeling of flows through complex regions via the~introduction of
explicit forcing terms in the~Navier--Stokes equations. Consistent
with the many shape-details of the~obstructing objects, such forcing
will need to represent the~perturbation of the~flow on various
length-scales simultaneously. This distinguishes the proposed
computational modeling from more conventional forced turbulence
procedures. In the latter the flow agitation is restricted to a~few
large scales only with the aim to observe the development of
a~natural inertial range at smaller scales in the turbulent flow
\cite{wang:examination:1996,kaneda:energy:2003}. Instead, in this
paper we~allow the~forcing of~a~collection of widely different
modes. The consequences for transport and dispersion in such
turbulent flows will be studied both in spectral - as well as in
physical space. We will primarily establish the degree by which
the~spectral properties of a turbulent flow can be modified relative
to the~classical Kolmogorov scaling, and~quantify the~efficiency
with which embedded scalar fields can be mixed by the modulated
flow.

Complementary to the proposed explicit forcing approach, two alternative formulations have been put forward in
literature to
capture the flow in and around complex objects. These include the
explicit boundary modeling \cite{whitaker:forchkeimer:1996} as well
as an approximation in terms of effective boundary conditions and
(surface) roughness parameters
\cite{hinze:turbulence:1975,jimenez:turbulent:2004}. The roughness
parameterization has been introduced for situations in which the
roughness length-scales are much smaller than the boundary layer
thickness \cite{townsend:structure:1976}. For geometries which
display both large- and small-scale contortions of the~shape of
the~object, compared to the~boundary layer thickness,
the~surface-roughness parametrization may not be sufficiently
accurate \cite{cheng:near:2002}. Alternatively, in~case of~explicit
boundary modeling, no-slip conditions are imposed at all the
intricate shape-details of the object. This computational approach
can in principle achieve full accuracy but is limited to cases of
modest complexity in view of the~elaborate geometric modeling and
the high computational expenses that are required
\cite{breugem:derivation:2005,breugem:direct:2005}.

Flow through complex gasket structures may give rise to self-similar
turbulence spectra which do not follow the well known Kolmogorov
$-5/3$ slope \cite{kolmogorov:1941}. Such non-Kolmogorov turbulence
was observed in flows over tree canopies, and~is reminiscent of
a~spectral short-cut feature that was also observed experimentally
\cite{finnigan:turbulence:2000}. In this paper we investigate
the~potential of multiscale forcing to accurately characterize such
dynamic flow-consequences of complex domain boundaries without the
need to explicitly account for their intricate geometrical shape.
We~consider the incompressible Navier--Stokes equations with
multiscale forcing working as a stirrer whose dynamical effects are
controlled by a distribution of simultaneously perturbed
length-scales.

To arrive at a~multiscale modeling that is quantitatively linked to
actual complex objects several steps need to be taken. In this paper
we~address a~first step in which we~examine in some detail
the~influence different forcing procedures have on the~energy
dynamics in spectral space and the mixing characteristics in
physical space. Special attention is devoted to the~mixing
efficiency of a~passive tracer by monitoring the~surface area and
wrinkling of level-sets of these~scalar fields
\cite{geurts:mixing:2001}. Specifically we~look at the~instantaneous
and accumulated effect on surface area and wrinkling caused by
broad-band forcing.

Different divergence-free forcing procedures will be applied to
directly perturb a large number of flow-scales. The alterations of
the flow dynamics express themselves clearly in the kinetic energy.
The transfer of energy toward smaller scales is found to increase
considerably, compared to the case in which only large scales are
forced. When a specific narrow band of scales is agitated by the
forcing, then the locally higher spectral energy is not `compatible'
with the molecular dissipation-rate and an accelerated transfer is
observed toward smaller scales. This effect is found for both
families of forcing methods, i.e., constant-energy and
constant-energy-input-rate. The~kinetic energy spectrum is also
modified non-locally, in a~range of scales that are larger than
the~directly forced scales. Consequently
the~agitation of a~band of small length-scale features can
accumulate and also induce significant alterations of the largest
flow-features, e.g., by contributing to an increased backscatter.

The~changes in the flow-dynamics due to the~application of
broad-band forcing application also has consequences for
the~turbulent transport properties of the~flow. This may be
expressed in terms of the~mixing efficiency of embedded passive
scalars. In~particular, monitoring the~surface-area of level-sets of
the~passive scalar allows to characterize changes in the~large-scale
sweeping of the flow, due to the~forcing. Likewise, the~more
localized motions directly affect the `wrinkling' of the~passive
scalar level-sets. The~dependence of these measures for the~mixing
efficiency on forcing parameters can be used to quantify the mixing
efficiency arising from agitation of different bands of
flow-structures with different forcing strengths. Specifically,
we~investigate the~dispersion of strongly localized initial scalar
concentrations. The~direct numerical simulation of the~forced
turbulence shows that the~maximal surface-area and wrinkling as well
as the~time at which such a~maximum is achieved can be controlled by
variation of forcing parameters. The~time-integrated surface-area
and wrinkling are indicators of the~accumulated effect.
The~simulations show that at small Schmidt numbers, a higher
emphasis on small-scale flow agitation yields a significant increase
in the time-integrated~total mixing of the~flow.

The~organization of this paper is as follows. In section
\ref{simmod} the simulation method, together with the explicit
forcing strategies are introduced. Section~\ref{modcas} is devoted
to the modulation of the cascading process associated with
the~different forcing methods. The consequences of forced turbulence
for transport and dispersion in physical space will be quantified in
section \ref{mixeff}. Concluding remarks are collected in section
\ref{concl}.

\section{Simulation of forced turbulence}
\label{simmod}

In this section we will first introduce the governing equations
(subsection \ref{goveq}) and subsequently describe the explicit
forcing strategies that are used to drive the flow (subsection
\ref{forc}). Two types of deterministic forcing strategies will be
included: procedures which yield constant-energy in the collection
of forced modes, and~procedures which correspond to a
constant-energy-input-rate for these modes. Subsequently, the
computational method, its validation and parallel performance will
be described (subsection \ref{compmeth}).

\subsection{Governing equations}
\label{goveq}

The dimensionless system of nonlinear partial differential equations
which governs the flow of a viscous incompressible fluid is given
by:
\begin{equation}\label{eq:ns}
\left\{ \begin{gathered}
  \frac{{\partial \mathbf{v}(\mathbf{x},t)}} {{\partial t}} +
  \Big( {\mathbf{v}(\mathbf{x},t) \cdot \nabla } \Big)\mathbf{v}(\mathbf{x},t)
  =  - \frac{1} {\rho }\nabla p(\mathbf{x},t) + \nu \nabla ^2 \mathbf{v}(\mathbf{x},t)
  + \mathbf{f}(\mathbf{x},t) \hfill \\
\nabla  \cdot {\mathbf{v}(\mathbf{x},t)} = 0 \hfill \\
\end{gathered}\right.
\end{equation}
where ${\mathbf{v}}$ is the velocity field, $\rho$ is the constant
mass-density and $p$ the pressure. The dimensionless viscosity is
the inverse of the computational Reynolds number
$\operatorname{Re}$, i.e., $\nu = 1/\operatorname{Re}$, and
$\mathbf{f}$ is the external forcing which we will specify in
subsection~\ref{forc}. This system of equations may be rewritten in
terms of the vorticity {\mbox{${\bf{\omega }}({\bf{x}},t) = \nabla
\times {\bf{v}}({\bf{x}},t)$}}. Making use of the~identity:
\begin{equation}\label{eq:identity1}
\Big( {\mathbf{v}(\mathbf{x},t) \cdot \nabla } \Big)\mathbf{v}(\mathbf{x},t) = \mathbf{\omega }(\mathbf{x},t) \times \mathbf{v}(\mathbf{x},t) +  \frac{1}{2} \nabla \Big( \left|{\mathbf{v}} (\mathbf{x},t) \right|^2\Big)
\end{equation}
we may express (\ref{eq:ns}) as:
\begin{equation}\label{eq:nsw}
\left( {\frac{\partial } {{\partial t}} - \nu \nabla ^2 } \right)\mathbf{v}(\mathbf{x},t) = \mathbf{w}(\mathbf{x},t) - \nabla \left( {\frac{1} {\rho } p(\mathbf{x},t) + \frac{1} {2}\left| {\mathbf{v}} (\mathbf{x},t) \right|^2} \right) + \mathbf{f}(\mathbf{x},t)
\end{equation}
where we introduced the nonlinear term $\mathbf{w}(\mathbf{x},t) =
\mathbf{v}(\mathbf{x},t) \times \mathbf{\omega}(\mathbf{x},t)$.

The~flow domain is assumed to be periodic with the same period in
each of the three coordinate directions. An efficient representation
of the solution in terms of Fourier modes can be adopted
\cite{canuto:spectral:1988,mccomb:physics:1990,young:investigation:1999}
in which the velocity ${\mathbf{v}}(\mathbf x, t)$ is expanded as:
\begin{equation} \label{eq:uk}
{\mathbf{v}} (\mathbf{x},t) = \sum\limits_\mathbf{k} {\mathbf{u}
(\mathbf{k},t)e^{\imath \mathbf{k} \cdot \mathbf{x}}}
\end{equation}
and the~wavevector $\mathbf{k}$ ($k=|\mathbf{k}|$) has components
$k_\alpha = 2 \pi n_\alpha/ L_b$, \mbox{$n_\alpha =0, \pm 1, \pm 2,
\ldots$} for $\alpha=1, 2, 3$. The dimensionless length of the
periodic domain is denoted by $L_b$ and $u_{\alpha}(\mathbf{k},t)$
is the~Fourier-coefficient corresponding to the~${\mathbf{k}}$-th
mode of $v_{\alpha}(\mathbf{x},t)$. The~equation governing
the~evolution of the~Fourier-coefficients is given by:
\begin{equation}\label{eq:nswk}
\left( {\frac{\partial }{{\partial t}} + \nu k^2 }
\right){\bf{u}}({\bf{k}},t) = {\bf{W}}({\bf{k}},t) - \imath {\bf{k}}
\mathcal F \left( {\frac{1}{\rho }p({\bf{x}},t) + \frac{1}{2}\left|
{\mathbf{v}} (\mathbf{x},t) \right|^2},{\mathbf{k}} \right) +
{\bf{F}}({\bf{k}},t)
\end{equation}
where ${\mathcal F}(a ({\mathbf{x}},t),{\mathbf{k}})$ denotes
the~Fourier-coefficient of the~function $a({\mathbf{x}},t)$
corresponding to wavevector ${\mathbf{k}}$:
\begin{equation}
{\mathcal F}(a ({\mathbf{x}},t),{\mathbf{k}}) =
A({\mathbf{k}},t)~~~{\mbox{if}}~~~a({\mathbf{x}},t) =
\sum_{{\mathbf{k}}} A({\mathbf{k}},t)e^{\imath \mathbf{k} \cdot
\mathbf{x}}
\end{equation}
In addition, ${\bf{W}}({\bf{k}},t)$ and ${\bf{F}}({\bf{k}},t)$
denote the~${\mathbf{k}}$-th Fourier-coefficient of the~nonlinearity
${\bf{w}}({\bf{x}},t)$ and forcing ${\bf{f}}({\bf{x}},t)$,
respectively.

In spectral space the~pressure term may be eliminated from
(\ref{eq:nswk}) if use is made of the~incompressibility condition.
This is equivalent to the well-known practice of solving a~Poisson
equation for the~pressure in physical space
formulations~\cite{wesseling:introduction:1992}. If we multiply
(\ref{eq:nswk}) by $\mathbf{k}$, use the continuity equation in
spectral space, i.e., $ {\bf{k}} \cdot {\bf{u}}({\bf{k}},t) = 0 $,
and assume that the forcing itself is divergence-free, so that $
{\bf{k}} \cdot {\bf{F}}({\bf{k}},t) = 0 $, the pressure term can be
written as:
\begin{equation}\label{eq:psk}
\mathcal F \Big( {\frac{1}{\rho }p({\bf{x}},t) + \frac{1}{2} \left|
{\mathbf{v}} (\mathbf{x},t) \right|^2}, {\mathbf{k}} \Big)=
\frac{{{\bf{k}} \cdot {\bf{W}}({\bf{k}},t)}}{{\imath k^2 }}
\end{equation}
The equation for the Fourier-coefficients of the velocity field
(\ref{eq:nswk}) may now be written as:
\begin{equation}\label{eq:nssW}
\left( {\frac{\partial }{{\partial t}} + \nu k^2 }
\right){\bf{u}}({\bf{k}},t) = {\bf{W}}({\bf{k}},t) -
{\bf{k}}\Big(\frac{{\bf{k}} \cdot {\bf{W}}({\bf{k}},t)}{k^{2}}\Big)
+ {\bf{F}}({\bf{k}},t)
\end{equation}
This may be expressed in a more compact form in terms of the
projection operator $\mathbf{D}$ defined by:
\begin{equation}
D_{\alpha \beta }  = \delta _{\alpha \beta }  - \frac{k_\alpha
k_\beta}{k^2}
\end{equation}
This operator restricts the solution to the~space of divergence-free
fields, represented by Fourier-coefficients
${\mathbf{u}}({\bf{k}},t)$ which lie in the~plane normal to
the~wavevector~$\mathbf{k}$. We~obtain the~governing equation for
the~desired Fourier-coefficients as:
\begin{equation}\label{eq:nssWD}
\left( {\frac{\partial }{{\partial t}} + \nu k^2 }
\right){\bf{u}}({\bf{k}},t) = {\bf{DW}}({\bf{k}},t) +
{\bf{F}}({\bf{k}},t)
\end{equation}
A~more detailed discussion of this spectral approach to
the~Navier--Stokes equations is available in
\cite{mccomb:physics:1990}. It forms the~basis for the~numerical
treatment that will be specified in subsection~\ref{compmeth}.

In various applications the dispersion of a passive scalar by
a~turbulent flow is of central importance. Passive scalar transport
may be used to characterize the~physical space consequences of
multiscale forced turbulence. The~governing equation for the
evolution of the scalar concentration $C({\bf{x}},t)$ contains
advection by the velocity field ${\mathbf{v}}({\bf{x}},t)$ as well
as diffusion. In physical space this may be expressed~as:
\begin{equation}\label{eq:psphys}
\frac{{\partial C(\mathbf{x},t)}} {{\partial t}} + \Big(
{\mathbf{v}(\mathbf{x},t) \cdot \nabla } \Big) C(\mathbf{x},t)
=\kappa \nabla ^2 C(\mathbf{x},t)
\end{equation}
where $\kappa$ is the non-dimensional molecular diffusivity of the
scalar. Compared to the dimensionless viscosity in (\ref{eq:ns}) we
adopt $\kappa=\nu/\operatorname{Sc}$ where the Schmidt number
$\operatorname{Sc}$ characterizes the scalar diffusion. Roughly
speaking, if ${\rm{Sc}} > 1$ then the scalar field displays a~wider
range of dynamically important length-scales, compared to
the~turbulent velocity field, while values ${\rm{Sc}} < 1$ indicate
a comparably smoother scalar field. The equation which governs
the~development of the~Fourier-coefficients $c({\mathbf{k}},t)$ of
the scalar field $C({\mathbf{x}},t)$ can readily be found~as:
\begin{equation}\label{eq:psspec}
\left( {\frac{\partial } {{\partial t}} + \kappa k^2 }
\right)c({\mathbf{k}},t) = Z({\mathbf{k}},t)~~~~~~{\rm{where}}~~~
Z({\mathbf{k}},t) = \mathcal{F} \Big({({\mathbf{v}}({\mathbf{x}},t)
\cdot \nabla) C({\mathbf{x}},t)},{\mathbf{k}} \Big)
\end{equation}
The~changes in the~turbulent transport properties of the~flow due to
the~multiscale forcing can be investigated by considering
the~evolution of the~scalar concentration at different Schmidt
numbers. The~structure of the~left-hand side of (\ref{eq:psspec}) is
identical to the~Navier--Stokes equations in (\ref{eq:nssWD}). This
allows to adopt the~same time-stepping method, as will be specified
in subsection~\ref{compmeth}.

To quantify the spectral-space effect of multiscale forcing, and
also to be able to concisely formulate the~different forcing
procedures in the next subsection, we~consider the~kinetic energy.
The equations which govern the~Fourier coefficients (\ref{eq:nssWD})
can be written in index notation as:
\begin{equation}\label{eq:nss}
  \left( {\frac{\partial } {{\partial t}} + \nu k^2 } \right)
  u_\alpha  (\mathbf{k},t) = \Psi_\alpha  ({\mathbf{k}},t) + F_\alpha
(\mathbf{k},t)
\end{equation}
where $\Psi _\alpha  ({\mathbf{k}},t)=  D_{\alpha \beta} W_\beta
(\mathbf{k},t)$ is the nonlinear term. Multiplying this equation by
the complex-conjugate $u_{\alpha}^{*}({\mathbf{k}},t)$ and summing
over the three coordinate directions, we obtain the kinetic energy
equation:
\begin{equation}\label{eq:eneeq0}
\left( {\frac{\partial } {{\partial t}} + 2\nu k^2 } \right)
E({\mathbf{k}},t) = {u_\alpha ^ * ({\mathbf{k}},t) \Psi _\alpha
({\mathbf{k}},t)} + {u_\alpha ^* ({\mathbf{k}},t) F_\alpha
({\mathbf{k}},t)}
\end{equation}
where $E({\mathbf{k}},t)= \frac{1}{2} \left|
{\mathbf{u}({\mathbf{k}},t)} \right|^2$ is the kinetic energy in
mode ${\mathbf{k}}$. Introducing the~notation for the~rate of energy
transfer $ T({\mathbf{k}},t) = {u_\alpha ^
* ({\mathbf{k}},t)\Psi _\alpha ({\mathbf{k}},t)}$, the~rate
of energy injection by the forcing $T_F ({\mathbf{k}},t) = {u_\alpha
^ * ({\mathbf{k}},t) F_\alpha ({\mathbf{k}},t)}$ and the~energy
dissipation rate $\varepsilon (\mathbf{k},t) = 2\nu k^2
E(\mathbf{k},t)$, we can write (\ref{eq:eneeq0}) equation as:
\begin{equation}\label{eq:eneeq1}
\frac{\partial E({\mathbf{k}},t)} {{\partial t}}  = - \varepsilon
(\mathbf{k},t) + T({\mathbf{k}},t) + T_F ({\mathbf{k}},t)
\end{equation}
This formulation clarifies that the rate of change of kinetic energy
$E(\mathbf{k},t)$ is connected with dissipation, expressed by the
viscous term $\varepsilon (\mathbf{k},t)$, with transfer to/from
different wavenumbers, expressed by $T(\mathbf{k},t)$, and~with
the~forcing term $T_F (\mathbf{k},t)$.

The~different contributions to the rate of change of the kinetic
energy typically act in distinct wavenumber regions. The
forcing~term $T_{F}(\mathbf{k},t)$ is non-zero in the forced modes
only. In this paper the collection of forced modes will always
contain a low wavenumber band corresponding to large-scale forcing
of the~flow. In addition, possible higher wavenumber contributions
can be included in $T_F(\mathbf{k},t)$. In contrast, energy
dissipation $\varepsilon (\mathbf{k},t)$ is defined in the~entire
spectral space, but it is dynamically important primarily for
the~high wavenumber range, i.e., acting on structures below
the~dissipation length-scale. Finally, the~transfer term
$T(\mathbf{k},t)$ is basic to the development of an~energy cascade
and is a dominant contribution for wavenumbers in an~inertial range
\cite{mccomb:physics:1990}. In the multiscale forcing cases, we will
also introduce forcing generally in the~same region as where
the~transfer $T(\mathbf{k},t)$ is dynamically important. Hence, the
effects of the multiscale forcing relate directly to
the~`competition' between the dynamics introduced by the forcing
procedure and the~`natural' transfer of energy to other modes in
the~spectrum.

In the~formulation of forcing procedures and in the~evaluation of
the kinetic energy dynamics, one frequently adopts shell-averaging.
The~basic operation consists of averaging over spherical shells of
thickness $2 \pi/ L_b$ centered around the origin. The~$n$-th
spherical shell is given by \mbox{$\frac{2\pi}{L_b}(n-1/2) <
|\mathbf{k}| \leq
 \frac{2\pi}{L_b}(n+1/2)$} and will be denoted by $\mathbb{K}_{n}$.
Applying shell-averaging to a function $h(\mathbf{k},t)$ defined in
spectral space we~obtain:
\begin{equation}\label{eq:shellavg}
{\overline{h}}(n ,t) = \frac{1}{P_{n}}\sum\limits_{\mathbb{K}_{n}}
{h({\bf{k}},t)}~~~;~~~P_{n}=\sum\limits_{\mathbb{K}_{n}} 1
\end{equation}
where $P_n$ is the number of modes in the $n$-th shell. Applying
the~shell-averaging (\ref{eq:shellavg}) to the energy equation
(\ref{eq:eneeq1}) we~end up with:
\begin{equation}\label{eq:eneeq2}
\frac{{\partial {\overline{E}}(n,t)}} {{\partial t}} =  -
{\overline{\varepsilon}} (n,t) + {\overline{T}}(n,t) +
{\overline{T}}_F (n,t)
\end{equation}
which indicates that the~interpretation of the various contributions
to the~rate of change of the~kinetic energy at mode ${\mathbf{k}}$
also applies to the shell-averaged formulation. In literature it is
common to introduce a numerical correction factor when averaging
over shells. This is used to compensate for the nonuniform
distribution of modes within the discrete spherical shells
\cite{eswaran:examination:1988,kerr:velocity:1990}. We will follow
the convention used in
\cite{eswaran:examination:1988,young:investigation:1999,yeung:response:1991}
when presenting the energy-spectra. This implies that we multiply
$\overline{h}(n,t)$ by a factor $4\pi n^2$ which is associated with
the `expected number modes' within the discrete shell. The
definition of the energy spectrum that we will adopt is given by
$E_{n} = \Big(4\pi n^{2}/{P_{n}}\Big)\sum\nolimits_{\mathbb{K}_{n}}
{E({\bf{k}},t)}$. Finally, summing (\ref{eq:eneeq1}) over all
wavevectors ${\mathbf{k}}$ or, equivalently, (\ref{eq:eneeq2}) over
all shells yields the evolution equation for the total energy in the
system:
\begin{equation}\label{eq:eneeq3}
\frac{{d {{\widehat E}}(t)}} {{d t}} =  - \widehat \varepsilon (t) +
\widehat T_F(t) ~~~;~~~ \widehat h(t)=\sum\nolimits_{n} {P_n
\overline{h}(n,t)} = \sum\nolimits_{{\mathbf{k}}}
{h({\mathbf{k}},t)}
\end{equation}
where use was made of the fact that the contribution of the transfer
term $T(\mathbf{k},t)$ is such that it only re-distributes energy
over the various modes, which implies that its sum over all
wavenumbers $\widehat T(t)=0$.

Next to spherical shells, it is convenient to introduce spherical
wavenumber bands which consist of several adjacent shells. We~denote
the~wavenumber band which consists of \mbox{$\frac{2\pi}{L_b}(m-1/2)
< |{\mathbf{k}}| \leq \frac{2\pi}{L_b}(p+1/2)$} by
$\mathbb{K}_{m,p}$, where \mbox{$m \leq p$}. The~corresponding
average over $\mathbb{K}_{m,p}$ of a~function $h({\mathbf{k}},t)$ is
given~by:
\begin{equation}\label{eq:bandavg}
\widetilde{h}^{(m,p)}(t) = \frac{1}{P_{m,p}} \sum_{n=m}^{p}
{P_n\overline{h}(n,t)} = \frac{1}{P_{m,p}} \sum\limits_{
\mathbb{K}_{m,p}} {h({\mathbf{k}},t)} ~~~;~~~ P_{m,p}
=\sum_{n=m}^{p} {P_n}
\end{equation}

To complete the computational model, we will next introduce the
explicit forcing strategies that will be investigated in this paper.

\subsection{Explicit forcing procedures}
\label{forc}

Forced turbulence in a~periodic box is one of the~most basic
numerically simulated turbulent flows. It is achieved by applying
large-scale forcing to the~Navier-Stokes equations. As a~result, at
sufficiently high Reynolds number the~well-known turbulent cascade
develops in an~inertial range of scales which are much smaller than
the~length-scale of the~forced modes
\cite{kolmogorov:1941,kolmogorov:refinment:1962,mccomb:physics:1990}.
The~statistical equilibrium that is reached is characterized by
a~balance between the input of energy through the~large-scale
forcing and the~viscous dissipation at scales beyond the~Kolmogorov
dissipation scale.

Various forcing procedures have been proposed in literature.
Generally, if~the~forcing is restricted to large scales only,
the~specific details of the~procedure do not have such a~large
effect on the properties of the developing inertial range at
sufficiently finer scales. However, since we wish to extend the
forcing to act on a~wide range of scales simultaneously, including
parts of an~inertial sub-range, the~differences between alternative
forcing procedures become more pronounced. Investigating these
differences is an~essential step toward quantitative modeling of
flow through complex gasket structures and forms the~main focus of
this paper. In this subsection, we~will recover the~definition and
some of the~motivation for several characteristic forcing
procedures.

\begin{figure}[hbt]
\centering{
\includegraphics[width=0.4\textwidth]{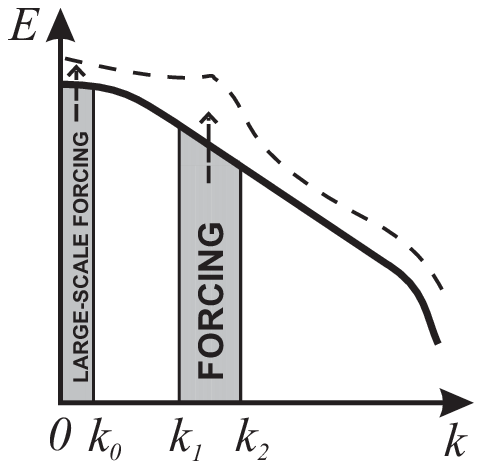}(a)
\includegraphics[width=0.3\textwidth]{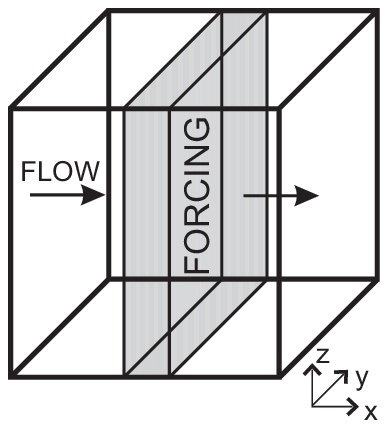}(b)
} \caption{Definition of two-band forcing in spectral space (a) and
localization of forcing within a~slab in physical space (b).}
\label{fig:foam}
\end{figure}

In~multiscale forcing, the~flow is agitated over a~wide range of
modes. To~investigate the~effects of such forcing we~will focus on
cases in which one additional spherical band of scales is forced,
next to the~common forcing of the~large scales. We~consider
the~general situation as depicted in Fig.~\ref{fig:foam}(a).
The~large scales are in the~range $k \leq k_0$ and an~additional
band of small scales is defined by $k_1 < k \leq  k_2$. The~forcing
method can also incorporate cases in which only part of the~domain
is occupied by a~complex obstruction, as sketched for the case of
a~slab in Fig.~\ref{fig:foam}(b). In~fact, by introducing
an~`indicator' function $\Theta ({\mathbf{x}},t)$ to locate
the~complex object within the~flow domain ($\Theta = 0$ outside
the~region occupied by the~object and $1$ elsewhere), the~forcing
can accommodate such spatial localization in a~flexible manner.
In~spectral space, the~introduction of $\Theta({\mathbf{x}},t)$
implies that the~forcing term in spectral space is represented by
the~convolution product of the~actual forcing $F(\mathbf{k},t)$ and
the Fourier-transform of the~indicator function. However,
in~the~present paper such complications will not be included and
we~will only consider forcing procedures applied in the~entire
physical domain.

Forcing procedures may be classified in different ways. We~first
distinguish forcing schemes which keep the total energy in
the~collection of forced modes identical to its value in the initial
condition. This will be referred to as class `A' forcing procedures.
Next, we identify forcing schemes which are characterized by
a~constant energy input rate, introduced via the collection of
forced modes. This group will be referred to as class `B'. In either
class of schemes, the~flows develop around a~well defined
statistically stationary state, but time-dependent variations in the
total energy and in the energy input rate may occur.

Apart from a~distinction concerning the way energy is introduced
into the~flow, one may classify forcing schemes as `deterministic'
or `stochastic'. Stochastic forcing schemes may introduce an~element
of uncorrelated randomness, e.g., by restricting the~forcing to
a~random subset of the~collection of forced modes every time
the~forcing is invoked. However, the primary question of locality of
the modulation of the~energy spectrum can be addressed more directly
using deterministic schemes and in this paper we will restrict to
these procedures. We~next introduce some characteristic forcing
schemes in either class `A' or class `B'.

\subsubsection*{Class `A': constant-energy forcing.~~~}

Various methods can be formulated which are such that the kinetic
energy in the forced modes remains constant. The simplest
possibility arises by requiring that $u_{\alpha}({\mathbf{k}},t)$
itself remains constant for all ${\mathbf{k}}$ in the collection of
forced modes. This was first proposed in
\cite{siggia:intermittency:1978} and implies for the forcing in
spectral space:
\begin{equation}\label{eq:siggia}
{\rm{A1}}:~~~ F_\alpha  ({\mathbf{k}},t) = \nu k^2 u_\alpha ({\mathbf{k}},t) - \Psi _\alpha  ({\mathbf{k}},t)
\end{equation}
One readily verifies, using (\ref{eq:nss}), that
$\partial_{t}u_{\alpha}({\mathbf{k}},t)=0$ and in particular this
implies that $\partial_{t}E({\mathbf{k}},t)=0$ for each of
the~forced modes. Hence, also the total kinetic energy contained in
all the~forced modes stays constant in time. The~energy input rate
corresponding to (\ref{eq:siggia}) is given by $T_F ({\mathbf{k}},t)
= \varepsilon({\mathbf{k}},t)- T({\mathbf{k}},t)$ for each of the
forced modes. This input rate may vary considerably in time, as the
unsteady flow will lead to a~strong time-dependence of the~energy
transfer $T$ for the~forced modes.

The~basic method (\ref{eq:siggia}) has motivated the formulation of
a number of extensions which relax the requirement that the
Fourier-coefficient is strictly constant. In
\cite{chasnov:simulation:1991} the method was modified to require
that  $\left| {\mathbf{u} ({\mathbf{k}},t)} \right| =
{\text{const}}$, i.e., equal to its initial value, for the forced
modes. This allows for the possibility that the phases of the
Fourier-coefficients may evolve in time. The~corresponding forcing
is given by:
\begin{equation}\label{eq:chasnov}
F_\alpha  ({\mathbf{k}},t) = \left( {\nu k^2  -
\frac{{T({\mathbf{k}},t)}} {{2E({\mathbf{k}},t)}}} \right)u_\alpha
({\mathbf{k}},t) = \left({{\varepsilon ({\mathbf{k}},t) -
T({\mathbf{k}},t)}} \right)  \frac{u_\alpha
({\mathbf{k}},t)}{{2E({\mathbf{k}},t)}}
\end{equation}
One may readily verify that this implies $\partial_{t}
E({\mathbf{k}},t)=0$ for the forced modes. Forcing expressed in
(\ref{eq:siggia}) or (\ref{eq:chasnov}) was found to yield quite
large fluctuations in the~energy input
rate~\cite{vincent:spatial:1991}.

Typically, the forced modes are
ordered according to the wavenumber shell to which they belong.
A~shell-oriented simplification of (\ref{eq:chasnov}) was
proposed~\cite{kerr:higher:1985,kerr:velocity:1990}:
\begin{equation}\label{eq:sullivan}
F_\alpha  (\mathbf{k},t) = \left( {\overline{\varepsilon}}(k,t) -
{\overline{T}}(k ,t) \right) \frac{u_\alpha (\mathbf{k},t)}
{2E(\mathbf{k},t)}
\end{equation}
This forcing also preserves the total kinetic energy in the forced
modes. The~three forcing procedures (\ref{eq:siggia}),
(\ref{eq:chasnov}) and (\ref{eq:sullivan}) are quite comparable,
both in terms of their fluid-physics motivation and in terms of
their turbulent flow predictions. Therefore, we will only present
actual simulation results obtained with (\ref{eq:siggia}), which are
quantitatively representative for the other two forcing procedures
in this group.

The~forcing methods described so far preserve the~kinetic energy
that is contained in the~collection of forced modes. However,
considerable variations in the~total energy in the~system can still
arise. The~reverse can also be realized, i.e., forced turbulence in
which the~total kinetic energy in the~system is constant, but
the~energy in different modes may vary in time. For this purpose,
the forcing should not be formulated in terms of quantities related
to individual modes or shell-averaged values, but rather contain
averages over all modes
\cite{grossman:scaling:1992,machiels:predictability:1997}. The case
of forcing in a single shell with $P$ modes can readily be
specified. Specifically, if we replace the shell-average
${\overline{(\cdot)}}$ in the amplitude factor in
(\ref{eq:sullivan}) by the average over all modes
${\widehat{(\cdot)}}$ and use the fact that ${\widehat{T}}=0$, we
obtain the forcing
\begin{equation} \label{eq:machiels}
{\rm{A2}}:~~~ F_\alpha  ({\mathbf{k}},t) = \frac{\widehat{\varepsilon}(t)}{P}
\frac{u_\alpha ({\mathbf{k}},t)} {2 {E}(\mathbf{k},t)}
\end{equation}
The~A2-forcing implies an~energy input rate
${\widehat{T}}_{F}={\widehat{\varepsilon}}(t)$ and thus by
(\ref{eq:eneeq3}) $d{\widehat{E}}/dt=0$. This method corresponds
exactly to the negative viscosity procedure used to maintain
quasi-steady turbulence direct numerical simulations results
reported in
\cite{jimenez:structure:1993,yamazaki:effects:2002,ishihara:high:2003,kaneda:energy:2003}.
Extension of A2-forcing to multiple shells can be realized in a
number of ways. This will be described in more detail momentarily.
A2-forcing will be compared to A1-forcing in the next section.

\subsubsection*{Class `B': constant-energy-input-rate forcing.~~~}

Next to forcing methods that can be associated with constant-energy,
one may define forcing procedures in which the total energy input
rate $\widehat T_{F}$ is constant. We~first present such forcing
methods with reference to a~single band of forced modes. The~way in
which the~energy input is distributed over several bands will be
specified afterwards.

A~central example in the class of constant-energy-input-rate forcing
methods was presented in \cite{ghosal:dynamic:1995}. Changing
$\widehat \varepsilon (t)$ in (\ref{eq:machiels}) into the constant
energy input-rate~$\varepsilon_{w}$, the corresponding forcing term
may be written as
\begin{equation}\label{eq:ghosal}
{\rm{B1}}:~~~ F_\alpha  ({\mathbf{k}},t) =  \frac{\varepsilon_w}{P}
\frac{u_\alpha ({\mathbf{k}},t)} {2 {E}(\mathbf{k},t)}
\end{equation}
The energy input-rate is found to be $\widehat T_F(t) =
\varepsilon_{w}$, as desired by construction. The total energy in
the system is no longer constant but governed by $d \widehat E(t)/dt
= -{{\widehat \varepsilon}}(t)+\varepsilon_{w}$ which implies that
the statistically stationary state that develops will show a
dissipation rate that fluctuates about $\varepsilon_{w}$. This type
of forcing was also studied in
\cite{misra:vortex:1997,carati:representation:1995,young:investigation:1999}.
Further extensions of the basic forcing procedure (\ref{eq:ghosal})
can be proposed in which an extra factor $k^{-q};~ q > 0$ arises in
the~definition of $F_{\alpha}$. Such an extra factor implies that
the~forcing of higher wavenumber shells can be made to correspond to
a~specific shape (usually $k^{-5/3}$ to more directly `impose'
Kolmogorov turbulence). These forcing procedures will not be
considered in this paper; for further details see
\cite{chen:high:1992,mohseni:numerical:2003}.

Similar to A-forcing methods, one may formulate related procedures
which are defined in terms of shell-averaged quantities. For
example, analogous to (\ref{eq:sullivan}), we may replace
$E({\mathbf{k}},t)$ in (\ref{eq:ghosal}) by ${\overline{E}}(n,t)$ to
define the forcing of modes in the $n$-th shell. This type of
forcing was found to yield basically the same results as those based
on (\ref{eq:ghosal}) and will not be presented explicitly in the
rest of this paper.

The final forcing procedure that we will include in this paper was
proposed recently in \cite{mazzi:fractal:2004}. It was motivated as
a~model of flow through a~fractal gasket which functions as
a~multiscale stirrer. This particular forcing may be associated with
a~constant energy input rate for the entire system. We~modify the~
original forcing procedure slightly and considered in particular
\begin{equation}\label{eq:mazzi}
{\rm{B}}2:~~~ F_\alpha  ({\mathbf{k}},t) = \frac{\varepsilon_{w}
k^\beta}{\sum_{ {\mathbf{k}} \in {\mathbb{K}} } \sqrt
{2E({\mathbf{k}},t)} k ^\beta } ~ e_\alpha  ({\mathbf{k}},t)
\end{equation}
where ${\mathbb{K}}$ denotes the set of forced modes. In this
formulation, the complexity of the object is parameterized by the
exponent $\beta$ which is related to the fractal dimension $D_{f}$
of the object through $\beta=D_{f}-2$. The vector
$\mathbf{e}(\mathbf{k},t)$ has the form:
\begin{equation} \label{eq:dirvect}
\mathbf{e}(\mathbf{k},t) =
\frac{\mathbf{u}(\mathbf{k},t)}{|\mathbf{u}(\mathbf{k},t)|} + \imath
\frac{\mathbf{k}\times \mathbf{u}(\mathbf{k},t)}{|\mathbf{k}|
|\mathbf{u}(\mathbf{k},t)|}
\end{equation}
which contains a part in the direction of ${\mathbf{u}}$ and a part
that is perpendicular~to~${\mathbf{u}}$. Since
$u^{*}_{\alpha}e_{\alpha}=|\mathbf{u}|=\sqrt{2 E({\mathbf{k}},t)}$
we find for the energy input rate:
\begin{equation}
T_{F}({\mathbf{k}},t)= \varepsilon_{w}  \frac{ k^\beta \sqrt{2
E({\mathbf{k}},t)}}{\sum_{ {\mathbf{k}} \in {\mathbb{K}} }  k ^\beta
\sqrt {2E({\mathbf{k}},t)} }
\end{equation}
In contrast to B1-forcing in which the energy input rate is constant
in time for each of the forced modes separately, this `fractal
forcing' procedure only implies a constant energy input rate for the
entire system. In fact, after summation over all forced modes the
total energy input rate is found to be equal to
${\widehat{T}}_{F}(t)=\varepsilon_{w}$. Correspondingly, we find for
the evolution of the total kinetic energy $d
{\widehat{E}}/dt=-{\widehat{\varepsilon}}(t)+\varepsilon_{w}$, i.e.,
identical as obtained before for B1-forcing. In the original
formulation in \cite{mazzi:fractal:2004} the energy input rate
$\varepsilon_{w}$ was replaced by the total dissipation rate
${\widehat{\varepsilon}}(t)$, which implies that ${\widehat{E}}$ is
constant in time.

So far, the B1- and B2-forcing methods were defined with reference
to a~single band of modes. This band was assumed to contain $P$
modes and~was identified by $\mathbb{K}$. The total energy input
rate $\varepsilon_{w}$ was available to this band. In case more
bands are forced simultaneously, the way the energy input-rate is
divided over the individual bands, and among the modes within each
band, needs to be specified. For two forced bands
$\mathbb{K}_{m_1,p_1}$ and $\mathbb{K}_{m_2,p_2}$ with $P_{m_1,p_1}$
and $P_{m_2,p_2}$ modes respectively, such a partitioning involves
two steps. First, a fraction $\varepsilon_{w,1}=a \varepsilon_{w}$
of the total energy input-rate is `allocated' to the first band and
the remainder $\varepsilon_{w,2}=(1-a)\varepsilon_{w}$ is used in
the forcing of the second band ($0 \leq a \leq 1$). Second, we
divide the energy input-rate that is available for each band equally
over all modes in the corresponding band. As an example, two-band
B1-forcing may be defined as
\begin{eqnarray}\label{eq:Pna}
{\rm{B1}}:~~~ F_\alpha  ({\mathbf{k}},t) &=& \frac{a\varepsilon_w}{P_{m_1,p_1}}\frac{u_\alpha ({\mathbf{k}},t)} {2 {E}(\mathbf{k},t)}~~~;~~~k \leq k_{0} \nonumber \\
&=&\frac{(1-a)\varepsilon_w}{P_{m_2,p_2}}\frac{u_\alpha ({\mathbf{k}},t)} {2 {E}(\mathbf{k},t)}~~~;~~~k_1 < k \leq k_2 \\
&=& 0 ~~~~~~~~~~~~;~~~~~~~~~\mbox{otherwise} \nonumber
\end{eqnarray}
The two-band formulation of B2-forcing can be specified analogously,
replacing $\varepsilon_{w}$ by either $a\varepsilon_{w}$ or
$(1-a)\varepsilon_{w}$ and $\mathbb{K}$ by $\mathbb{K}_{m_1,p_1}$ or
$\mathbb{K}_{m_2,p_2}$, respectively. Extending A2-forcing to more
bands can be done in a similar way in which a~fraction $a
\widehat{\varepsilon}(t)$ is associated with the~large-scale band
and the~remainder with the~second band. {The~specific choice of
$P_{m_1,p_1}$ and $P_{m_2,p_2}$ above implies that the~energy is
equally distributed between all modes within a~forced band. We can
go one step further and require the equal distribution of
$\varepsilon_{w}$ over the forced shells contained in the bands.
This implies changing $P_{m_1,p_1}$ and $P_{m_2,p_2}$ into the
number of modes $P_{n}$ for each forced shell.} Extension to more
forced bands can be formulated analogously.

In~the~next subsection the numerical method will be specified in
some detail, to complete the~description of the physical and
computational modeling used in this paper.

\subsection{Computational method and parallelization}
\label{compmeth}

In this subsection we first specify the time-stepping method, then
sketch some aspects of the implementation and subsequently describe
the~validation of~the~method.

\subsubsection*{Time evolution.~~~}

To simulate the spectral solution governed by equations
(\ref{eq:nssWD}) and (\ref{eq:psspec}) we~first rewrite these
equations in a~more general form having in mind that the~evolution
due to the diffusive terms can be computed exactly by introducing
integrating factors $e^{\nu k^2 t}$ and $e^{\kappa k^2 t}$,
respectively \cite{canuto:spectral:1988}. In~fact, (\ref{eq:nssWD})
and (\ref{eq:psspec}) may be expressed as:
\begin{equation} \label{eq:rk}
\frac{{\partial \mathbf{U}(\mathbf{k},t)}} {{\partial t}} =
\mathbf{G} \left( {\mathbf{U}(\mathbf{k},t)} \right)
\end{equation}
where
\begin{eqnarray*}
{\bf{U}} =  \left[
              {\begin{array}{*{20}c}
                {{\mathbf{u}} ({\bf{k}},t)e^{\nu k^2 t}}  \\
                {c({\bf{k}},t)e^{\kappa k^2 t}}  \\
               \end{array}}
            \right]~~~;~~~
{\bf{G}}  = \left[
              {\begin{array}{*{20}c}
                {\left( {\mathbf{D}} {\mathbf{W}}  ({\bf{k}},t) + {\mathbf{F}}  ({\bf{k}},t) \right) e^{\nu k^2 t}} \\
                {Z ({\bf{k}},t)e^{\kappa k^2 t}}  \\
               \end{array}}
            \right]
\end{eqnarray*}
We~use a~constant time-step $\Delta t$ to obtain the~solution at
times \mbox{$t_{n} = t_0 + n \Delta t$}. A four-stage, second-order,
compact-storage Runge-Kutta method was implemented. The advancement
of the solution over a~full time-step requires four steps of
the~form:
\begin{equation}
\mathbf{U}(\mathbf{k},t_{n + \gamma} ) = \mathbf{U}(\mathbf{k},t_n )
+ \gamma \Delta t~\mathbf{G} (\mathbf{U}(\mathbf{k},t_{n+\xi}))
\end{equation}
The~intermediate solutions in the~different stages can be found as
follows. In~stage 1 we~adopt $(\gamma,\xi)=(1/4, 0)$, stage 2
requires $(\gamma,\xi)=(1/3, 1/4)$, stage 3 uses $(\gamma,\xi)=(1/2,
1/3)$ and stage 4 completes the~step with $(\gamma,\xi)=(1,
1/2)$~\cite{geurts:modern:2001}.

We~consider turbulence in a~cubic box of side $L_b$ with periodic
boundary conditions and assume that the flow is statistically
isotropic which implies that we require the same resolution in each
coordinate direction. The direct numerical simulations will employ
a~resolution of $N^3$, where $N$ is the number of spectral-space
grid-points that is used in each direction. This restricts the~set
of wavenumbers to $n_\alpha =0, \pm 1, \pm 2, \ldots, \pm \left(
{N/2 - 1} \right), - N/2$. The~cut-off wavenumber is given by
{\mbox{$k_{\max} = \pi N/ L_b$.}} In physical space this corresponds
to a~uniform grid \mbox{$x_\alpha = j L_b/N $,} where $ j = 0, 1,
2,..., N - 1$ in each coordinate direction.

We~use the~pseudo-spectral discretization method, i.e., the~spatial
derivative terms in the~Navier--Stokes and passive scalar equations
are computed via simple multiplications in the~spectral space.
The~nonlinear terms in the~equations are evaluated in physical space
to avoid the evaluation of several computationally intensive
convolution sums \cite{canuto:spectral:1988}. This procedure
requires three steps. First, the Fourier-coefficients
${\mathbf{u}}({\mathbf{k}},t)$ and $c({\mathbf{k}},t)$ are used to
obtain the~velocity and scalar fields in the~physical space.
Subsequently, the~velocity-velocity products and the~velocity-scalar
products are determined in physical space and finally the associated
Fourier-coefficients of these products are obtained.

The~finite resolution may give rise to well-known aliasing errors.
In fact, the~product of two Fourier-series based on a~resolution
with $N$ points gives rise to more small-scale modes
than can be supported by the~grid. As~a~result, these contributions
can appear on the~$N$-point resolution as seemingly lower
wave-number modes. A~detailed discussion of techniques allowing
the~partial or full removal of the aliasing error can be found in
\cite{canuto:spectral:1988}. {Several of these methods were
tested, as described in the appendix, closely
following~\cite{rogallo:illiac:1977,rogallo:numerical:1981}.}
The~method of two shifted grids and spherical truncation was used in
actual simulations. This removes the aliasing error completely which
was found to be essential, especially to maintain
the~characteristics of the small turbulent scales.

\subsubsection*{Data decomposition and fast Fourier transforms.~~~}

The~simulation software was implemented in Fortran 90 and
parallelized based on the framework given in
\cite{young:investigation:1999} using the Message Passing Interface
(MPI) \cite{MPI}. Data are stored using the Hierarchical Data Format
(HDF5) \cite{HDF5} which is a~file format and library designed for
scientific data-storage and handling. The choice of HDF5 was
motivated by the~flexible data-exchange between different platforms
and its support of parallel~I/O. High performance computations were
done at SARA Computing and Networking Services (Amsterdam) on
Silicon Graphics (SGI) Altix 3700 and Origin 3800 CC-NUMA systems
(for more details see \cite{SARA}).

The~critical performance factors in the~parallel implementation of
the~pseudo-spectral discretization method are the~domain
decomposition and~the~algorithm for the~three-dimensional Fast
Fourier Transform (FFT). These two implementation decisions are
essential since they determine almost all aspects of
the~data-exchange between domains and most of the~floating point
operations. It is important to obtain a~data decomposition which
permits for fast transfer of data between processors. To obtain
parallel Fourier transforms we adopted procedures from two
libraries: SCSL \cite{scsl} and FFTW \cite{FFTW98}. Moreover, since
access to memory and the~number and speed of available \mbox{CPU-s} may
differ considerably among different computational platforms,
significant improvements in the~processing time can be achieved by
platform-dependent optimization.

The~speedup of the parallel implementation was checked by
simulating~decaying turbulence at a resolution of $256^3$.
A~time-interval $0 \leq t \leq 0.05$ was considered. This case
corresponds to $28$ time-steps with $5$ data evaluation and
reporting stages. In~a~non-dedicated SGI~Altix 3700 environment
we~obtained on $4, 8, 16, 32, 64$ processors the~following speedup
numbers: $3.9, 7.5, 14, 26, 47$, respectively. The~best performance
results were obtained by a~cache-unfriendly parallelization along
the second array dimension. This gives the~opportunity of minimal
data exchange and reshuffling between processors and illustrates
that the~speed of the processors overwhelms the abilities of direct
access to the~memory. This was found to be the critical issue for
the hardware that was available.

\subsubsection*{Code validation.~~~}

To validate the~implementation of the~pseudo-spectral method,
decaying homogeneous isotropic turbulence was simulated at two
different Reynolds numbers. The initial condition was taken from
\cite{meyers:database:2003}, which was generated on the~basis of
the~Pao spectrum \cite{pope:turbulent:2000}. For further details
we~refer to \cite{meyers:accuracy:2004}. This flow was studied
extensively using high-order finite-volume discretization and
explicit Runge-Kutta time-stepping. Special attention was given to
the~degree of convergence that could be achieved using the finite
volume approach. These data provide a clear point of reference with
which the~present pseudo-spectral flow-solver can be compared.

\begin{table}[htb]
\begin{center}
\begin{tabular}{c|c|c|c|c|c|c|c|c|c|}
$R_\lambda / N^3$ & $32^3$ & $48^3$& $64^3$& $96^3$& $128^3$& $192^3$& $256^3$ & $384^3$ & $512^3$ \\
\hline$50 $ & $0.56$ & $0.83$& $\underline{\mathbf{1.11}}$& $1.67$& $2.22$&
$3.34$& $4.45$ & $6.67$ & $8.90$
\\ \hline $100 $ & $0.20$ & $0.29$& $0.39$& $0.59$& $0.79$&
$\underline{\mathbf{1.18}}$& $1.57$ & $2.36$ & $3.15$   \\
\end{tabular}
\caption{The value of $k_{\max} \eta$ associated with  different
resolutions. The~Kolmogorov scales are \mbox{$\eta =
5.87\cdot10^{-3}$} and~\mbox{$\eta= 2.07\cdot10^{-3}$} for
$R_\lambda=50$ and $R_\lambda=100$, respectively.}
\label{tab:resolution}
\end{center}
\end{table}

A~first, global assessment of the resolution that is achieved may be
inferred by evaluating the product of the cut-off wavenumber and the
observed Kolmogorov dissipation length-scale $\eta= L(3R_\lambda ^2
/20)^{ - 3/4}$ in terms of the Taylor Reynolds number $R_{\lambda}$
computed for the initial condition (see (\ref{eq:erls}) for the
definition) and integral length $L=1/2$. In order to resolve all
dynamically relevant length-scales, including the dissipation
length-scale it is required that $k_{\max}\eta$ is sufficiently
large. A commonly accepted criterion of adequate spatial resolution
is that $k_{\max}\eta > 1$. When the focus is on higher-order
statistics, it is preferred to use larger values ($k_{\max}\eta >
3/2)$ \cite{wang:examination:1996,eswaran:examination:1988}.
In~Table \ref{tab:resolution} the values of $k_{\max} \eta$ are
presented for the two computational Reynolds numbers considered
$Re=1060.7$ and $Re=4242.6$ which correspond to $R_{\lambda}=50$ and
$R_{\lambda}=100$. We~observe that in the~first case a resolution of
at least $64^{3}$ is required to achieve full resolution, while in
the second case the minimal required resolution moves up to
$192^{3}$.

For validation of the code, the~flow was simulated for more than two
eddy-turnover times and a~number of quantities were monitored:
\begin{eqnarray}
{\rm{Total~energy:}}~~~&& \widehat E(t)=\sum \nolimits_{\mathbf{k}} {{E}(\mathbf{k},t)} \\
{\rm{Taylor~microscale:}}~~~&& \lambda(t)=\left( 5\widehat E(t)/\sum \nolimits_{\mathbf{k}} {k^{2}{E}(\mathbf{k},t)} \right)^{1/2} \\
{\rm{Taylor~Reynolds:}}~~~&& R_\lambda (t)=\lambda(t)u(t)/ \nu~~~;~~~u(t)=\sqrt{\frac23 \widehat E(t)} \label{eq:erls}\\
{\rm{Longitudinal~skewness:}}~~~&& S_1(t)=-\frac{{\left\langle
(\partial {v_1(\mathbf{x},t)}/\partial {x_1})^3\right\rangle}}{
{\left\langle (\partial {v_1(\mathbf{x},t)}/\partial
{x_1})^2\right\rangle^{3/2}}}  \label{eq:erlss}
\end{eqnarray}
The~operator $\left\langle \cdot \right\rangle$ in (\ref{eq:erlss})
refers to volume averaging.

\begin{figure}[hbt]
\centering{
\includegraphics[width=0.45\textwidth]{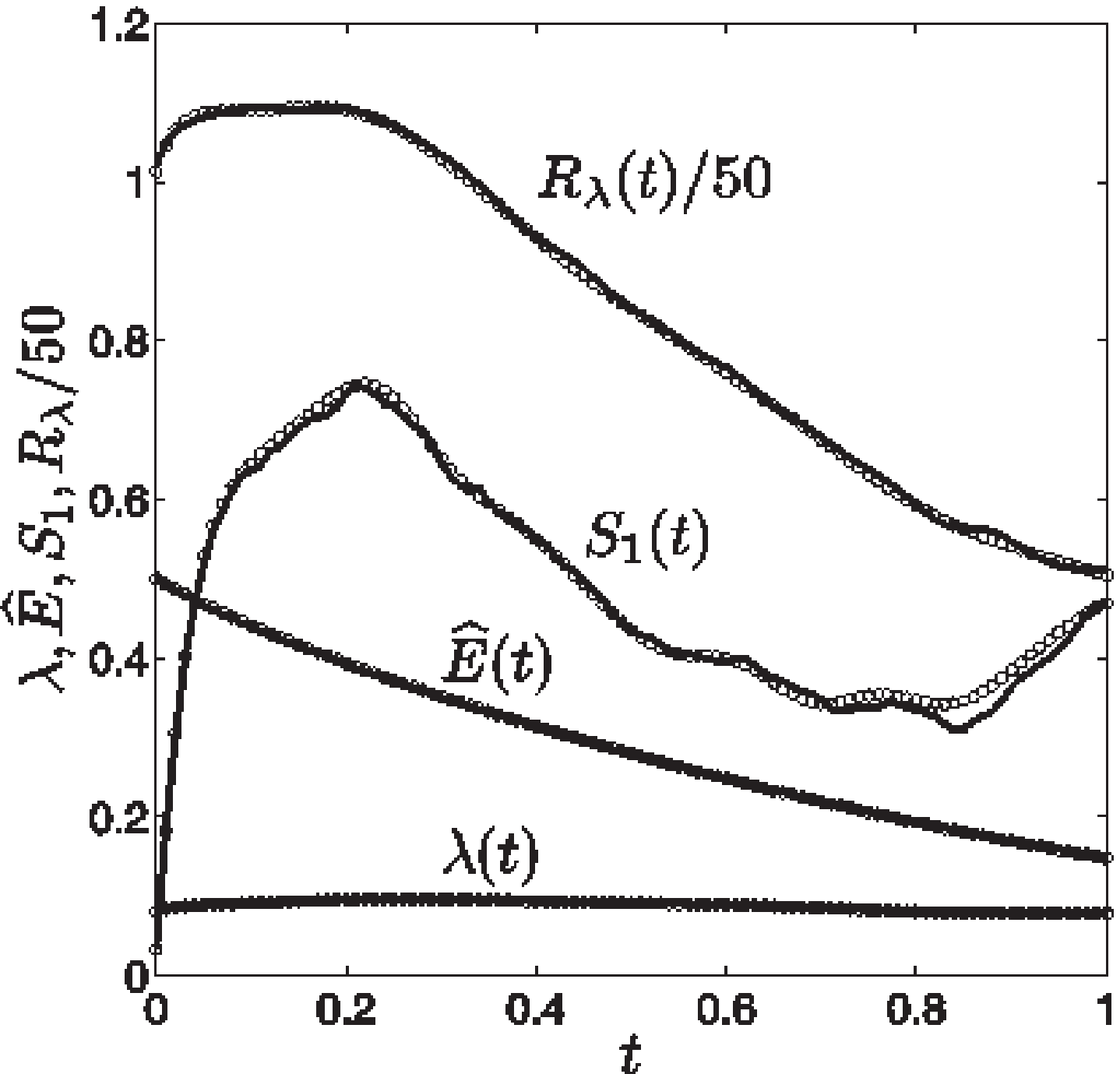}(a)
\includegraphics[width=0.45\textwidth]{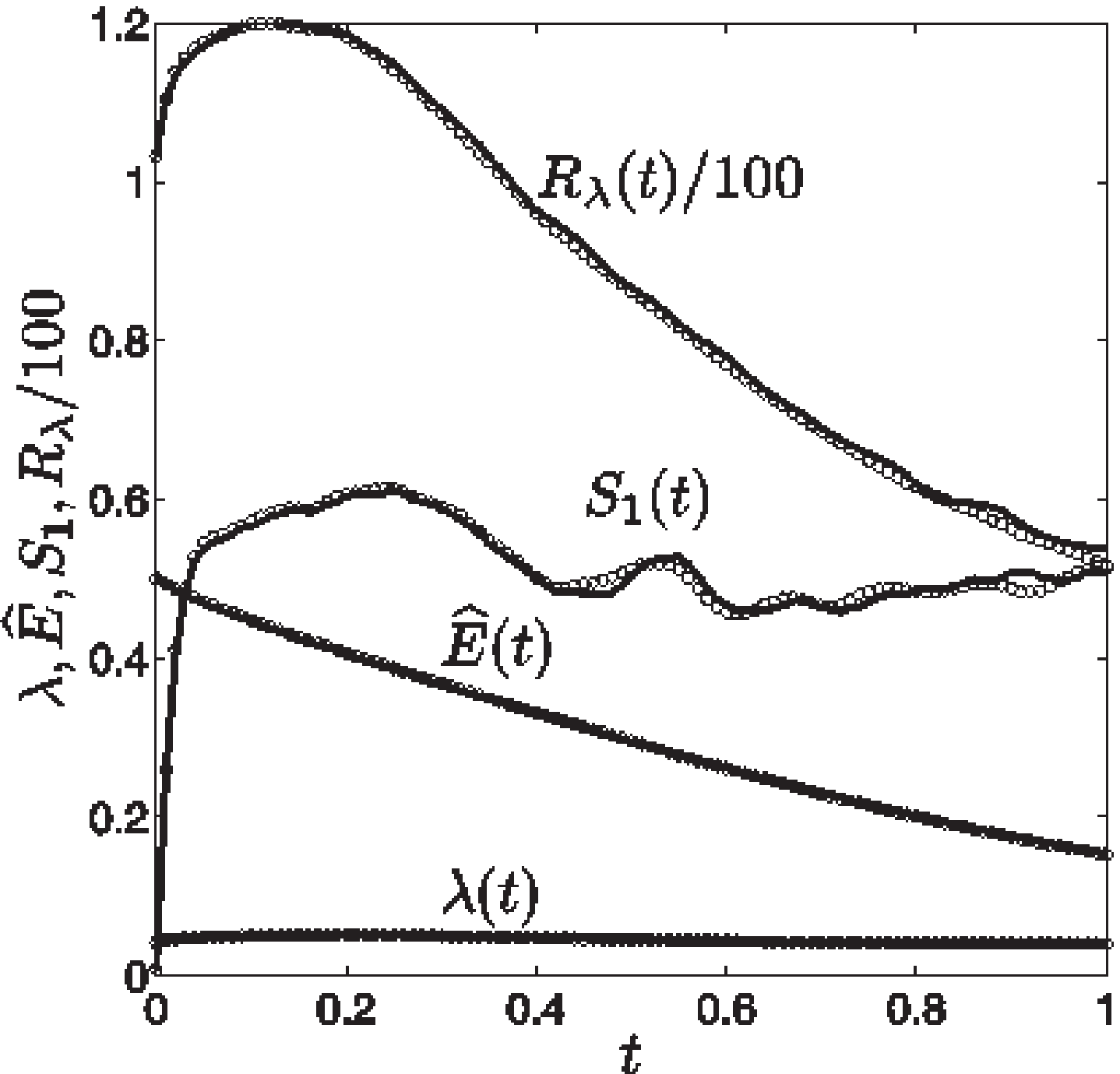}(b)}
\caption{Prediction of total energy $\widehat{E}$, Taylor microscale
$\lambda$, Taylor Reynolds number $R_{\lambda}$ and longitudinal
skewness $S_{1}$ at an initial $R_{\lambda}=50$ (a) and
$R_{\lambda}=100$ (b) with a finite-volume
\cite{meyers:database:2003}~(solid) and the present pseudo-spectral (dotted) code.}
\label{fig:validation1}
\end{figure}

In Fig.~\ref{fig:validation1}, a~comparison is made between
simulation results obtained with the~pseudo-spectral method at
$N=512$, and with the high-order finite-volume discretization method
\cite{meyers:database:2003}. For each of the~quantities an~almost
perfect agreement may be observed.
\begin{figure}[hbt]
\centering{
\includegraphics[width=0.45\textwidth]{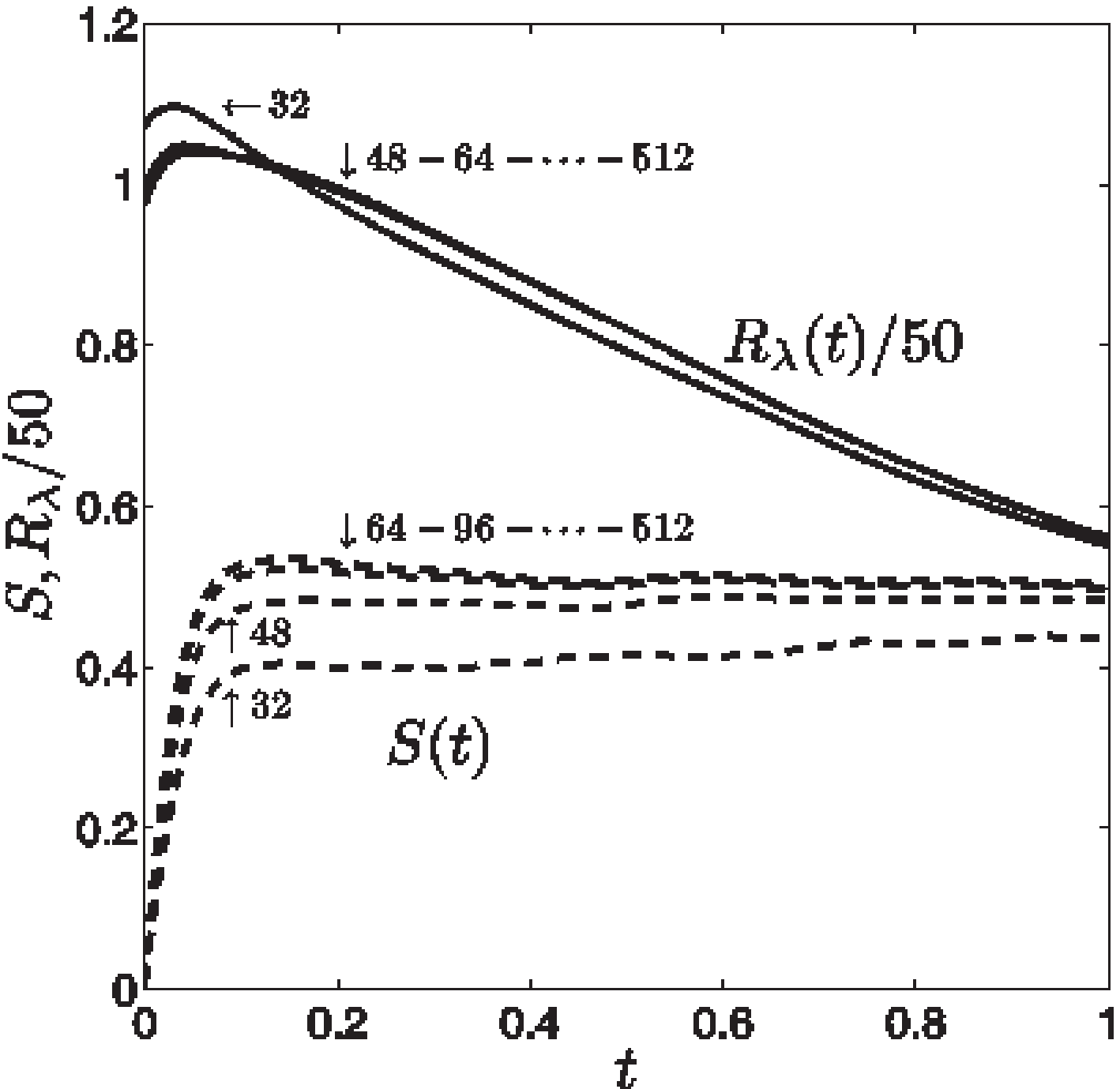}(a)
\includegraphics[width=0.45\textwidth]{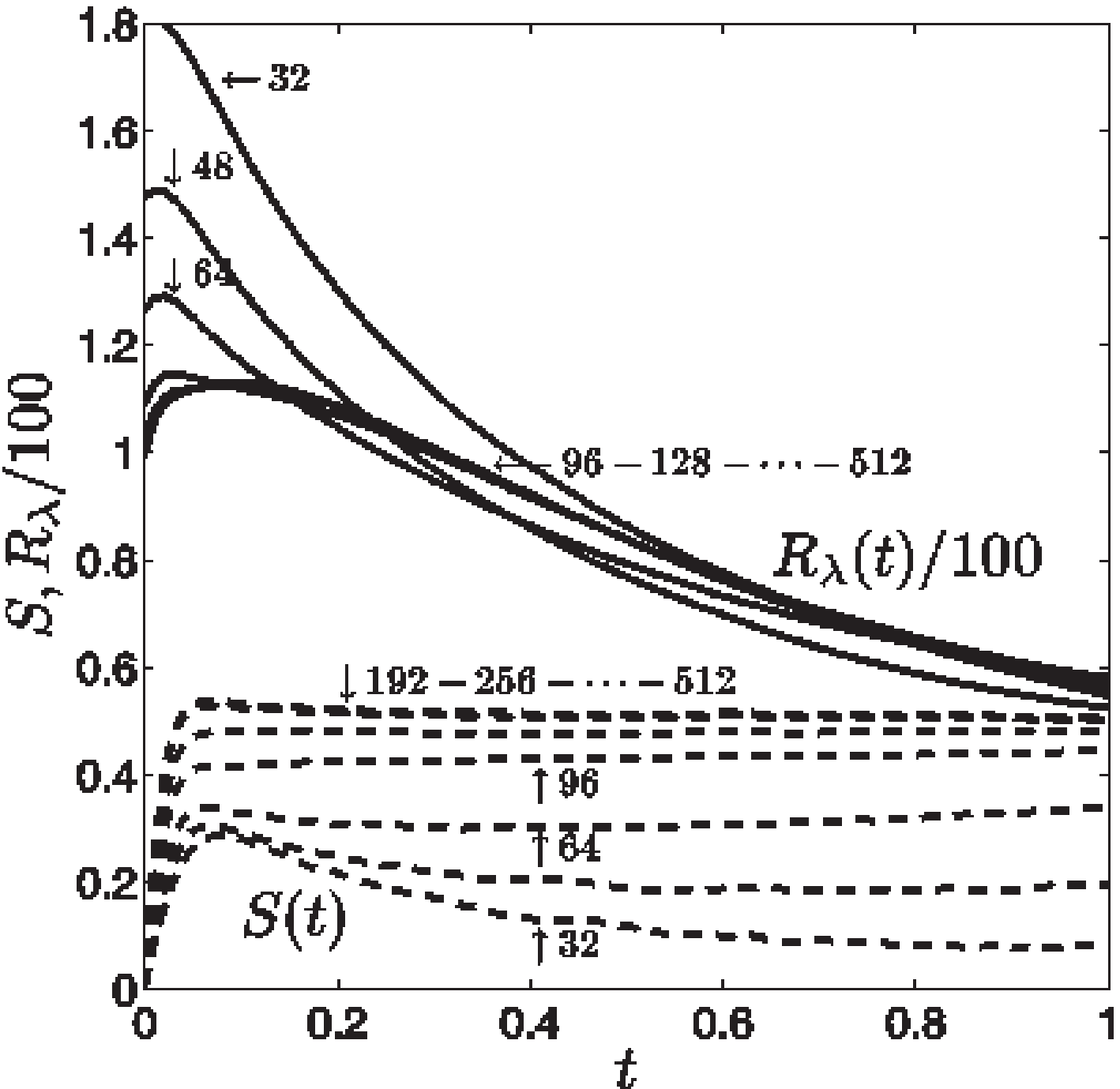}(b)}
\caption{Convergence of predictions for the Taylor-Reynolds number
$R_{\lambda}$ and the skewness $S$ at an initial $R_{\lambda}=50$
(a) and $R_{\lambda}=100$ (b) for different resolutions
$N=32\cdot48\cdot64\cdot96\cdot128\cdot192\cdot256\cdot384\cdot512$.}
\label{fig:validation2}
\end{figure}
In~Fig.~\ref{fig:validation2} we~assessed the~convergence of
the~predictions as function of the~spatial resolution. In~this
figure we~replaced the~longitudinal skewness $S_{1}$ by
the~skewness:
\begin{equation}
S(t) = \frac{2} {{35}}\left( {\frac{{\lambda (t)}} {{u(t)}}}
\right)^3 \sum\nolimits_{\mathbf{k}} {k^2 {{T}}(\mathbf{k},t)}
\end{equation}
For homogeneous isotropic turbulence the value of $S$ should be
equal to~$0.5$~\cite{batchelor:theory:1953},which is quite well
approximated in the simulations. This quantity is quite sensitive to
the~spatial resolution and is therefore a~good indicator of
appropriate spatial resolution. We~observe that the~different
predictions display a clearly distinguishable convergence toward the
grid-independent solution. Specifically, results obtained for
resolutions higher than $64^3$ at $R_{\lambda}=50$ and $192^3$ at
$R_{\lambda}=100$ are quite indistinguishable, consistent with
the~criterion that $k_{\max}\eta
> 1$.

In~the~next section we~turn to the~effects that different multiscale
forcing procedures have on the~developing turbulent flow. We~will
focus in particular on the~modifications that arise in the~kinetic
energy spectrum.

\section{Modulated cascading by broad-band forcing}
\label{modcas}

The~explicit forcing in different wavenumber bands can have a~strong
effect on the developing turbulent flow. We~discuss the
modifications of the~energy spectrum arising from `constant-energy'
{\mbox{(class `A')}} or `constant-energy-input-rate' (class `B')
procedures. The~various forcing strategies will be shown to
qualitatively correspond to each other, provided the total
dissipation-rate $\varepsilon_{w}$ and the spectral energy
distribution are commensurate for the different class `A' and `B'
forcing strategies. We will specify this inter-relation in more
detail momentarily. As point of reference, we will first turn our
attention to forcing of the large scales only. Subsequently, we
consider two-band forcing and investigate in particular the effects
of variation of the strength and location of the small-scales band
on the developing flow.

In the sequel, we~consider time-averaged properties of
the~developing turbulent flow defined~by:
\begin{equation}\label{eq:ht}
\left\langle {h} \right\rangle_t  = \lim_{t\rightarrow \infty}
\frac{1}{{t - t_0 }}\int\limits_{t_0 }^t {{h(\tau)} d\tau} \approx
\frac{1}{{{\cal{T}} - t_0 }}\int\limits_{t_0 }^{\cal{T}} {{h(\tau)}
d\tau}
\end{equation}
where ${\cal{T}}$ is sufficiently large. In all cases $t_0=5$ in
order to allow the averaging-process to start from a properly
developed quasi-stationary state. The averaging is continued up to
${\cal{T}}=25$, which corresponds to approximately 40 eddy-turnover
times. This was found to provide an accurate representation of
the~long-time averages, leading to relative errors below
$5$~percent, measured in terms of the ratio of the standard
deviation and the~mean signal. This procedure was applied to obtain
the~time-averaged kinetic energy spectra as well, which are very
effective for monitoring changes in the~kinetic energy dynamics due
to the forcing.

\subsubsection*{Large-scale forcing.~~~}

To~create a~point of reference, we~first consider forced turbulence
in which energy is introduced to the~system only in the first
shell~$\mathbb{K}_{1,1}$. We adopt $k_{0}=3\pi$ referring to
Fig.~\ref{fig:foam}(a) and force all $18$ modes inside this band.
The~computational Reynolds number $Re=1060.7$ and the~size of the
computational domain $L_b=1$. The spatial resolution was taken to be
$128^{3}$, which provides ample resolution of these cases, similar
to what was established in subsection~\ref{compmeth}.

In order to be able to quantitatively compare results obtained with
the different forcing strategies, care should be taken of properly
`assigning' a level for the energy dissipation-rate and the spectral
energy distribution. For this purpose, we may consider simulations
with the A2-method to be central in the sense that the other three
forcing strategies may be specified with reference to it. In fact,
if we generate an initial condition with a certain total kinetic
energy, then A2-forcing yields an evolving flow which becomes
statistically stationary after some time, while maintaining the same
level of total energy. The A2-forced simulation can be used to
specify the `corresponding' class-B forcing strategies. In fact, the
constant dissipation-rate $\varepsilon_{w}$ in class `B' forcing is
taken equal to the time-average value of the dissipation-rate that
is found from the A2-forced simulation, i.e., we adopt
$\varepsilon_{w}=\langle \widehat{\varepsilon} \rangle_{t}$. This
procedure was adhered to in all cases presented in this section.
Finally, in the developed stages of either these A2- or B-forced
flows, any instantaneous solution may be used to arrive at a full
specification of the `corresponding' A1-forcing. The actual choice
of this instantaneous solution is arbitrary. However, when comparing
simulations based on A1-forcing that adopt different realizations of
the turbulent flow-field, we observed that the statistical
properties of all these A1-forced cases were the same.

\begin{figure}[hbt]
\centering{
\includegraphics[width=0.45\textwidth]{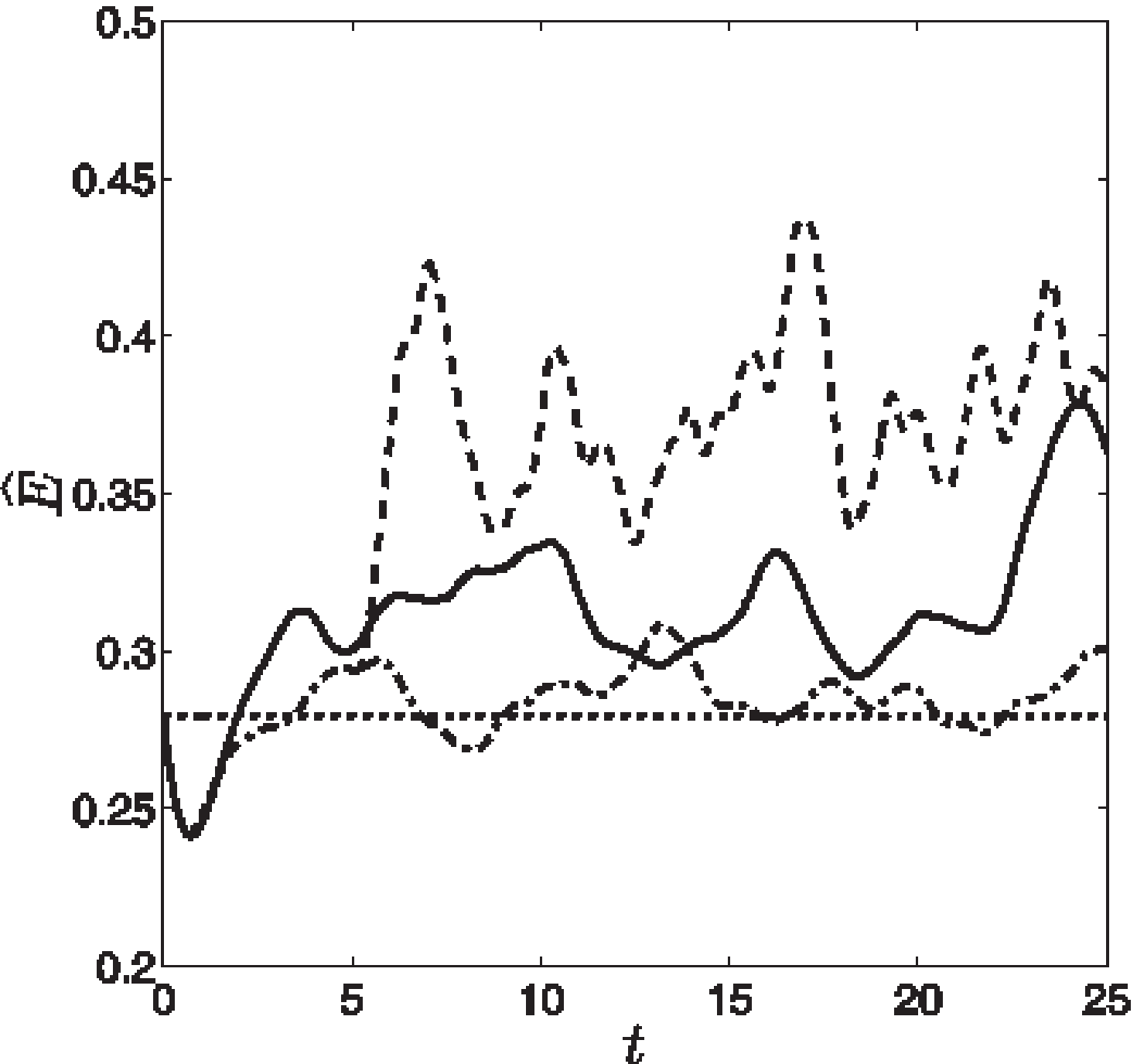}(a)
\includegraphics[width=0.45\textwidth]{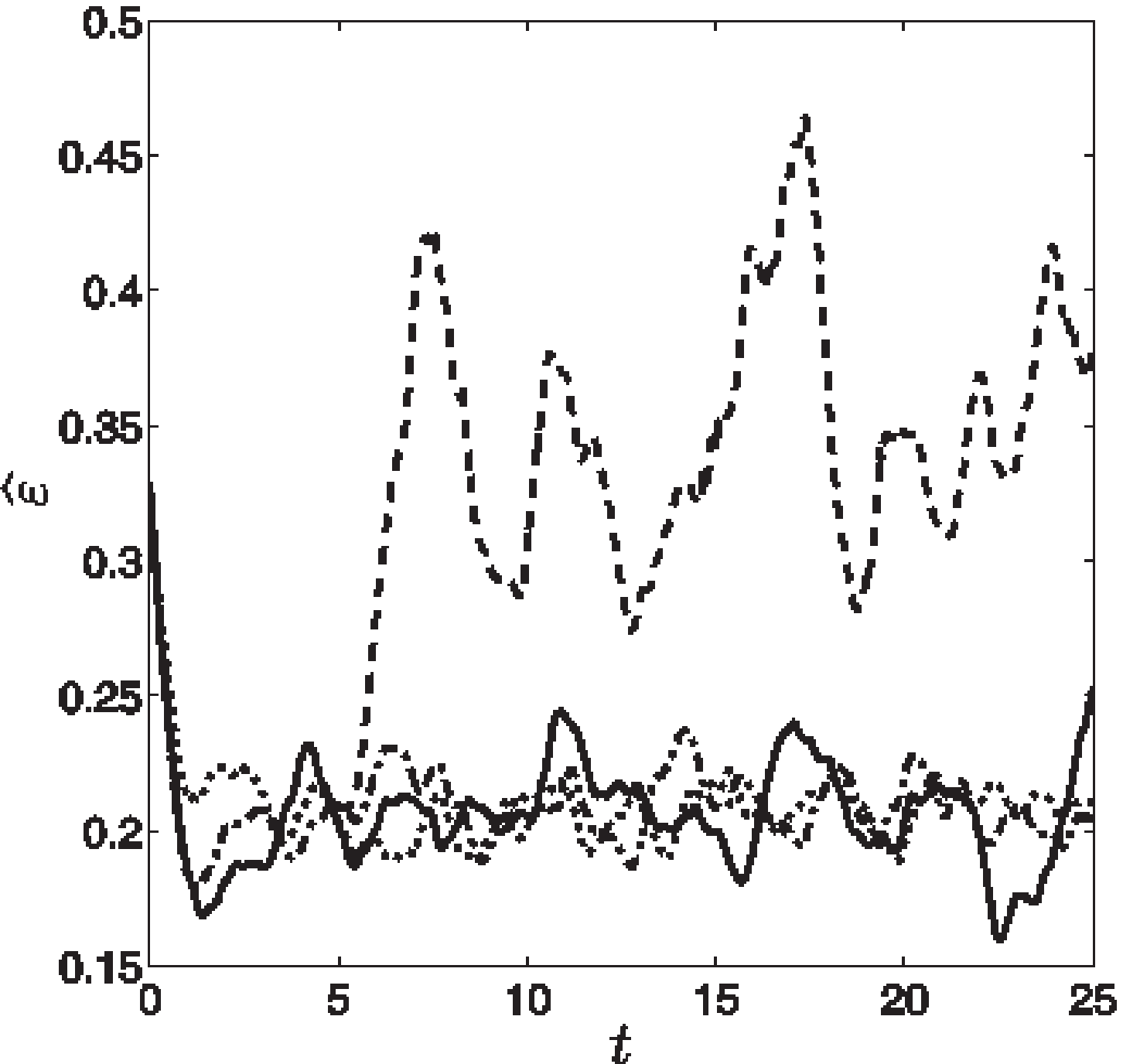}(b) \\
} \caption{The evolution of the total kinetic energy
$\widehat{E}$~(a) and energy-dissipation-rate
$\widehat{\varepsilon}$~(b) for the~large-scale forcing:
A1~(dashed), A2~(dotted), B1~(dash-dotted), B2~(solid).}
\label{fig:stat0}
\end{figure}

The~evolution of the~total kinetic energy~$\widehat{E}(t)$ and
energy dissipation rate~$\widehat{\varepsilon}(t)$ is shown in
Fig.~\ref{fig:stat0}. As initial condition for the A2- and B-forced
simulations, we adopted the velocity field obtained at $t=0.5$ from
the decaying homogeneous turbulence simulation discussed in
subsection~\ref{compmeth}. To be able to qualitatively compare with
the A1-forced flow at a similar energetic level we took as~initial
condition the solution from B1-forcing  at $t=5$. The~total kinetic
energy is seen to fluctuate around its long-time mean value (of course, apart
from A2-forcing). As
can be seen, the~system rapidly develops into a statistically
stationary state characterized by the input of energy, its transfer
to smaller scales and dissipation in the viscous range.
In~A1-forcing the Fourier-coefficients in the forced band are all
kept constant, i.e., equal to their~initial values.
The energy in the~system fluctuates very significantly, which was
considered a disadvantage of this forcing in
\cite{eswaran:examination:1988}. The~energy and dissipation levels
in A1-forcing differ considerably from those obtained with the other
forcing strategies. To compare A2-forcing with B-forcing, the~energy
dissipation rate was taken as $\varepsilon_w = \left\langle
{\varepsilon} \right\rangle_t \cong 0.2$. The~total kinetic energy
for B1-forcing is seen to fluctuate around the constant value
associated with A2-forcing. A~similar impression is observed when
use is made of the fractal B2-forcing  in which the~fractal
dimension of the stirrer was taken equal to $D_{f}=2.6$
\cite{mazzi:fractal:2004} which corresponds to an exponent
$\beta=3/5$ in (\ref{eq:mazzi}).

In~general, when applied to the~largest scales only, all forcing
procedures mentioned in subsection~\ref{forc} yield similar results.
As~a~further example, the~tails of the time-averaged spectra were
found to be virtually identical to each other, which indicates that
the properties of the~smaller turbulent length-scales are not very
strongly dependent on the~details of the~specific forcing. This was
also established by various other quantities that were investigated.
Specifically, the Taylor-Reynolds number $R_\lambda $ for the
simulated cases was seen to fluctuate in the range between $\approx
50$ up to $\approx 60$ for all methods. The~time-averaged value of
the~skewness was also investigated and found to be very close to
$0.5$. This indicates that a~well developed isotropic flow was
attained~\cite{batchelor:theory:1953}.

\subsubsection*{Two-band forcing.~~~}

In~the~simulations that adopt two-band forcing we~consider
situations in which we introduce energy into the~system in a~band
consisting of four shells, next to the already described large-scale
forcing in the first shell. We~first compare the different class `A'
and `B' forcing strategies, within this two-band setting. As second,
forced band we consider $\mathbb{K}_{17,20}$. This band corresponds
to $k_{1}=33\pi$ and $k_{2}=41\pi$ in Fig.~\ref{fig:foam}(a) and
contains in total $17284$ different modes that are all explicitly
forced. The~comparison of the~different forcing strategies shows
that the~flow-predictions are qualitatively comparable.
Subsequently, we~therefore focus on the~B2-forcing strategy and
investigate the~effects arising from changes in the~strength or
the~location of the~second forced band.

Forcing of a second band implies that we need to additionally
specify how the energy input is distributed over the bands,
the~shells within the bands and, finally, the~modes within the
shells. The~specification of the A2-forcing requires the~fraction of
the energy-input that is allocated to the different bands.
We~consider the case in which $a=1/5$ in (\ref{eq:Pna}) which
corresponds to equi-partitioning of the energy-input over the five
shells that are forced. The~forcing within the~second band is
further specified by assigning an equal energy-input rate to each of
the~four shells contained in it. Finally, each of the modes in
a~particular shell $n$ receives an equal share of the energy-input
to that shell, taking the number $P_{n}$ of modes in the particular
shell into account. To compare the `A' with `B' forcing strategies
we adopt the~same method as above for specifying the parameter
$\varepsilon_{w}$. Specifically, the total energy injection for the
`B' methods was given as  $\varepsilon_w =\langle
\widehat{\varepsilon} \rangle_{t} \cong 1$, in terms of the
time-average of the total dissipation rate in the~A2-forcing.
Moreover, the same equi-partitioning of the energy-input as in
A2-forcing was adopted. Finally, the A1-forcing is derived from the
field that was obtained at $t=5$ with the B1-forcing. We verified
that the statistical properties of the A1-forced flow are
insensitive to the particular choice of initial field used to define
this forcing method.

\begin{figure}[hbt]
\centering{
\includegraphics[width=0.45\textwidth]{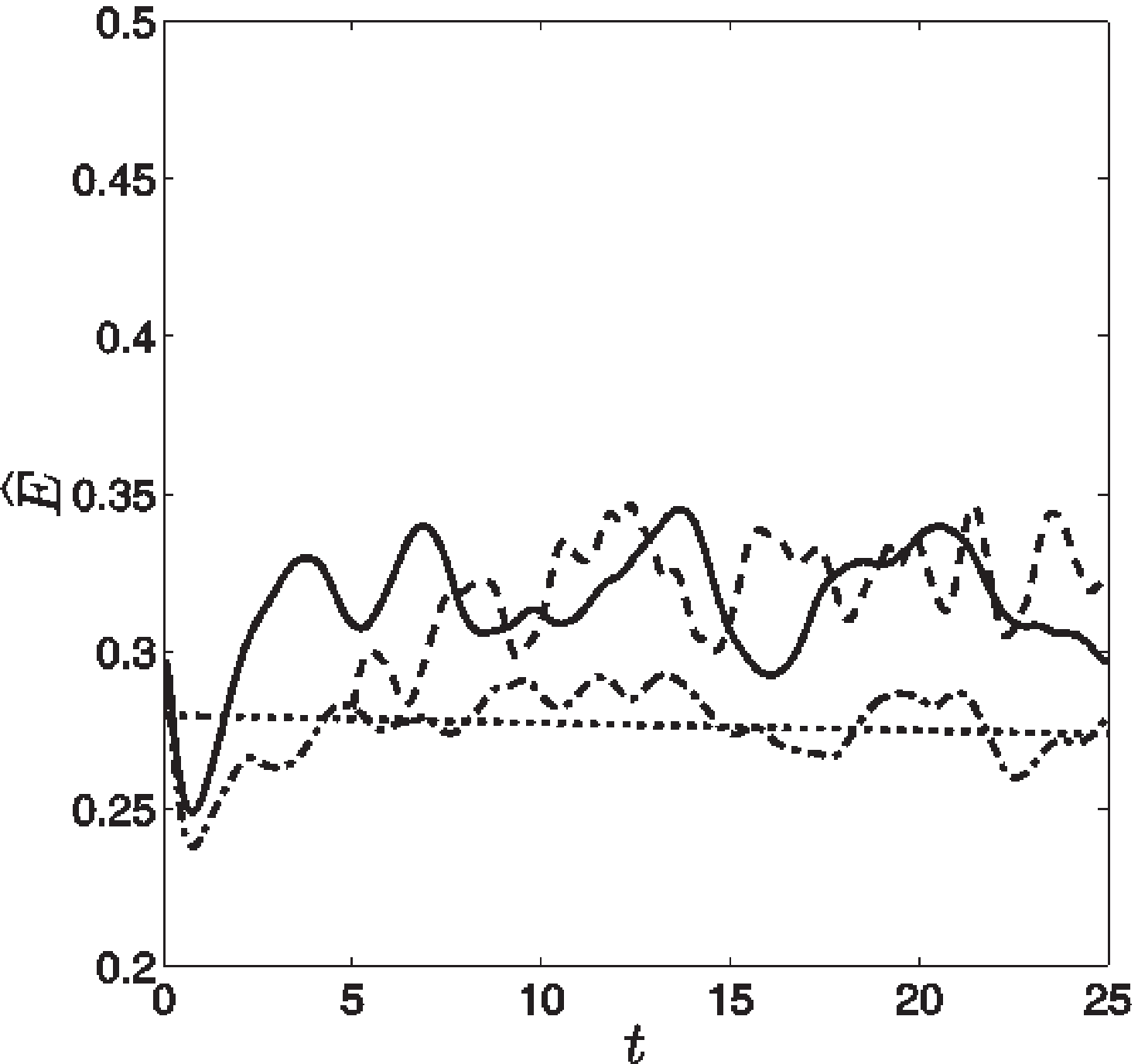}(a)
\includegraphics[width=0.45\textwidth]{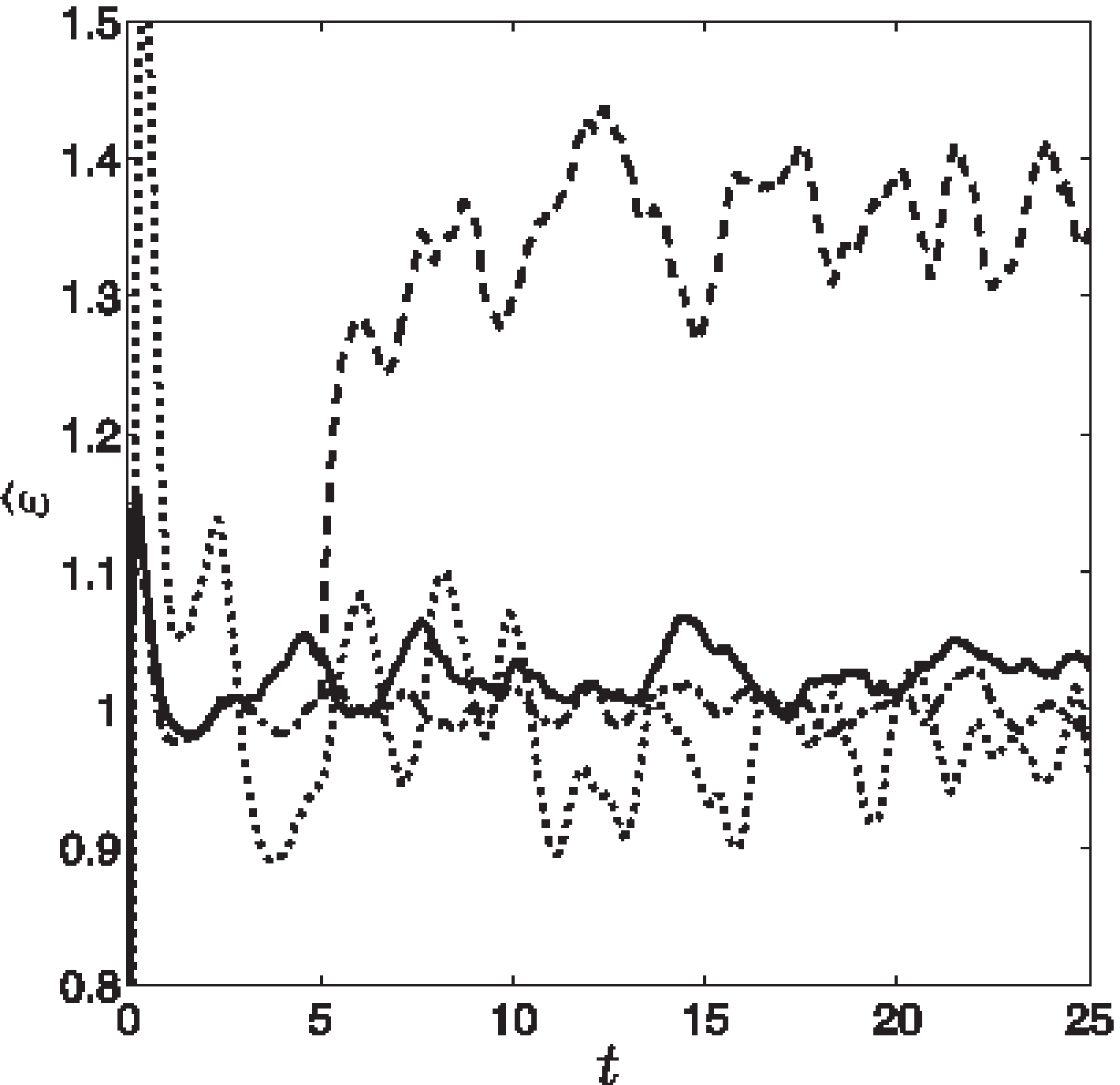}(b)
} \caption{The evolution of the total kinetic energy
$\widehat{E}$~(a) and energy-dissipation-rate
$\widehat{\varepsilon}$~(b) for two-band forcing: A1~(dashed),
A2~(dotted), B1~(dash-dotted), B2~(solid).} \label{fig:stat1}
\end{figure}

As may be noticed by comparing Fig.~\ref{fig:stat0} with
Fig.~\ref{fig:stat1}, the two-band forcing leads to a strong
increase in the total energy dissipation-rate, while the~total
kinetic energy present in the flow is quite unaffected by the second
forced band. The increase in the dissipation-rate is particularly
strong for A1-forcing. Hence, the high-$k$ forcing changes mainly
the {\it{distribution}} of energy over the~scales and not so much
the actual energy content. By changing the~strength and location of
the~forcing, we~have the~possibility to control, and to some extent
manipulate the~way the~energy is distributed and hence indirectly
influence the~large- and small-scale transport properties of
the~flow. We~turn to this aspect next, by focusing explicitly on
the~kinetic energy spectrum.

\begin{figure}[hbt]
\centering{
\includegraphics[width=0.45\textwidth]{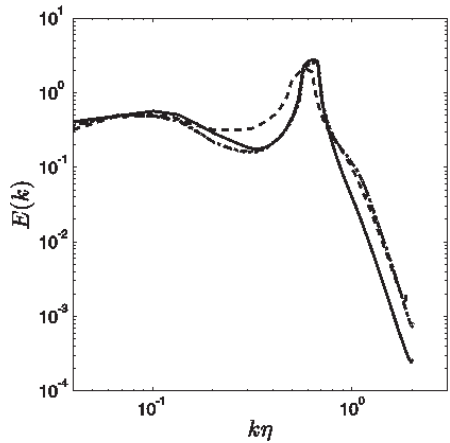}
} \caption{Compensated energy spectrum for two-band forced
turbulence ($k \leq 3\pi$ and $33\pi < k \leq 41\pi$) with different
methods: A1~(dashed), A2~(dotted), B1~(dash-dotted), B2~(solid).}
\label{fig:spectra_2bands}
\end{figure}

The compensated kinetic energy spectra  $E(k)=\left\langle {\widehat
\varepsilon} \right\rangle_t^{-2/3} k^{5/3} \left\langle
E_k\right\rangle_t$ that are obtained with the different two-band
forcing methods are collected in Fig.~\ref{fig:spectra_2bands}.
The~modifications in the~spectrum, relative to the case of
large-scale forcing only, are localized primarily in the~region
close to the~forced band. All forcing methods are seen to yield
qualitatively quite similar results. Next to the expected
modifications near the explicitly forced band, we observe that the
two-band forcing also affects a~much wider set of larger scale
modes. In fact, a significant depletion of the~kinetic energy in a
range of scales `ahead of' the forced region, is readily
appreciated. This indicates that the~agitation of a~small band of
modes can induce large changes in a~rather wide part of the~spectrum
which further characterizes the type of turbulence-control that one
may achieve with explicit forcing.

\begin{figure}[hbt]
\centering{
\includegraphics[width=0.45\textwidth]{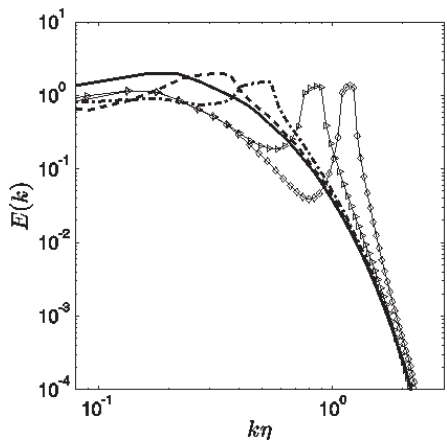}
} \caption{Compensated energy spectrum for two-band forced
turbulence with the B2-method and the same energy inputs
$\varepsilon_{w,1}=\varepsilon_{w,2}=0.15$ to the $k \leq 3\pi$ band and
various locations of the~second band: $9\pi < k \leq 17\pi$, $17\pi
< k \leq 23\pi$, $33\pi < k \leq 41\pi$, $49\pi < k \leq 57\pi$
(dashed, dash-dotted, $\triangleright$, $\diamond$). The spectrum obtained with large-scale
forcing at $\varepsilon_{w}=0.15$ in $k \leq
3\pi$ band (solid).} \label{fig:spectra_2bands_b}
\end{figure}

We next turn to the second part of this section and consider the
effects of varying the spectral support and the strength of the
second forced band. The qualitative similarity of the different
two-band forcing methods as seen in Fig.~\ref{fig:spectra_2bands}
allows to concentrate on only one of the forcing methods. We~adopt
B2-forcing in the sequel. In Fig.~\ref{fig:spectra_2bands_b} we
illustrate the effect of variation of the spectral support of
the~second band. Relative to the~case of large-scale forcing only,
we~observe that the~tails of the~spectra are quite unaffected. However,
the~injection of energy in the~second band is seen not only to
increase the~energy in the~forced scales but also to deplete
the~energy in all the~larger scales. Moreover, the~`up-scale' effect
of energy-depletion is more pronounced in case the second band is
moved toward smaller scales.

\begin{figure}
\begin{center}
\includegraphics[width=0.45\textwidth]{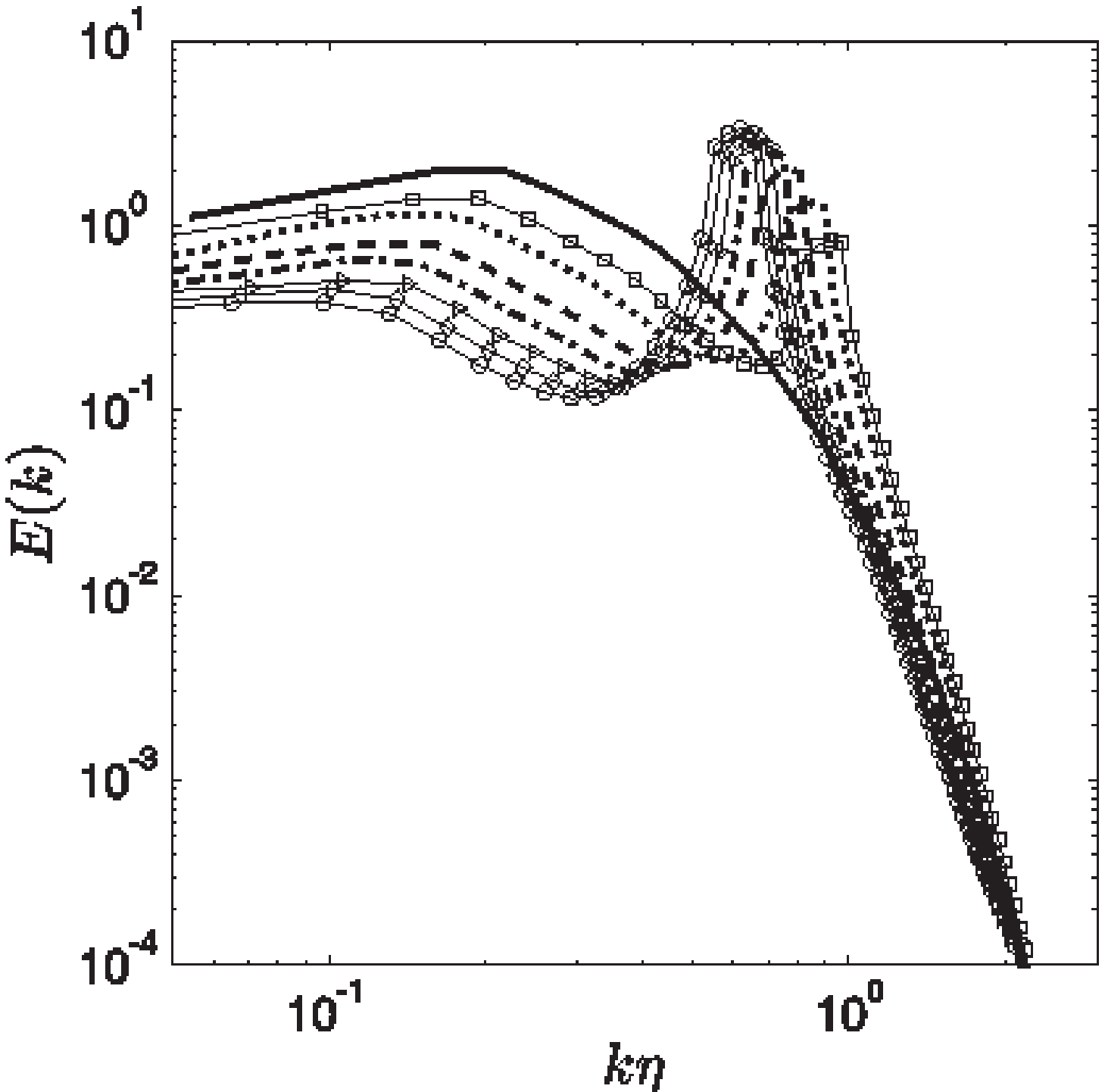}(a)
\includegraphics[width=0.45\textwidth]{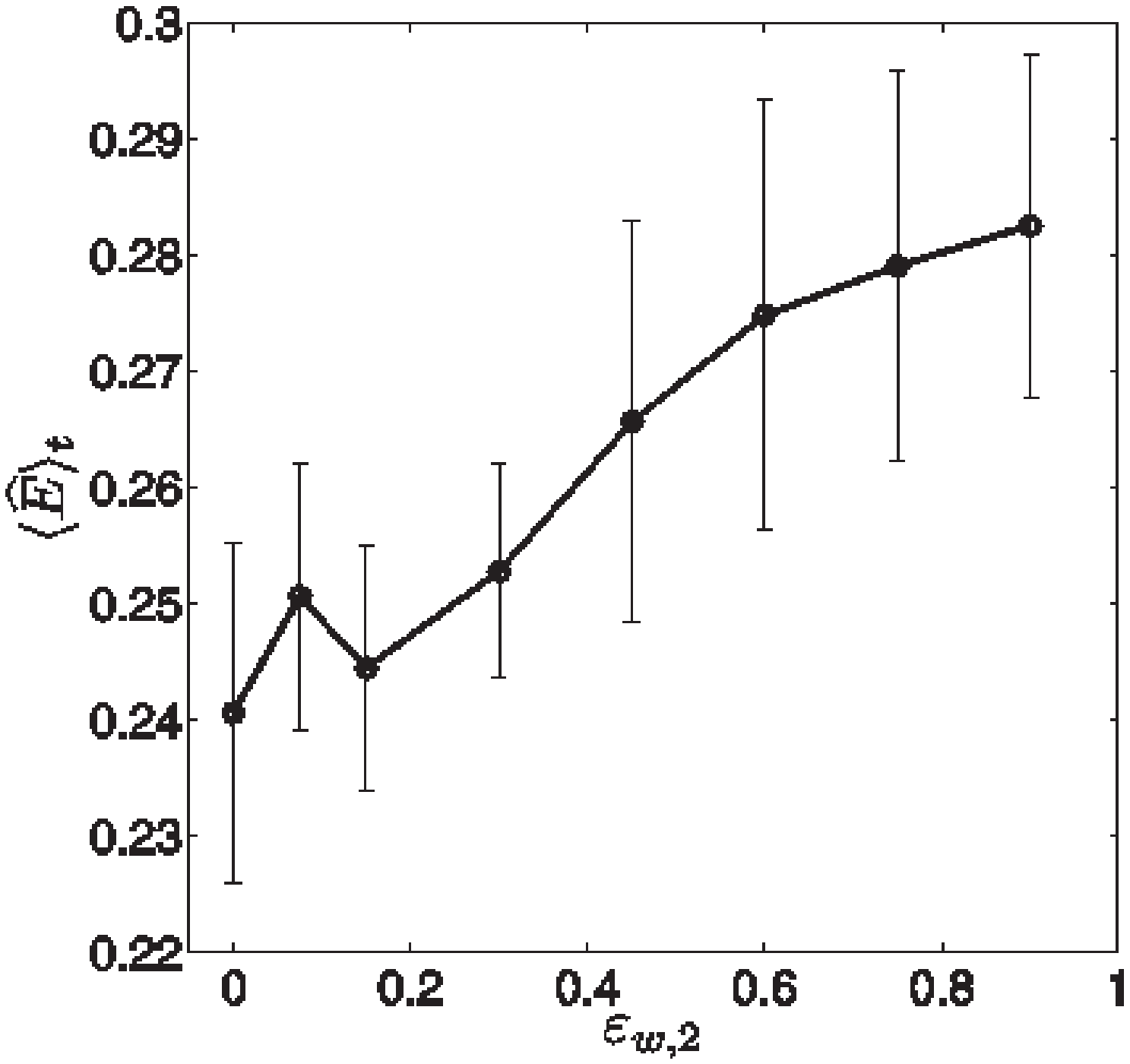}(b)
\end{center}
\caption{(a) Compensated energy spectrum for two-band forced
turbulence with the B2-forcing method and energy input rate
$\varepsilon_{w,1}=0.15$ in $k \leq 3\pi$ band for different
strengths of forcing in the $33\pi < k \leq 41\pi$ band:
$\varepsilon_{w,2} = 0.075, 0.15, 0.30, 0.45, 0.60, 0.75, 0.90$
($\square$, dotted, dashed, dash-dotted, $\triangleright$,
$\diamond$, $\circ$). Large-scale forcing with
$\varepsilon_{w}=0.15$ in the $k \leq 3\pi$ band is denoted by the solid line.
(b) Corresponding time-averaged total kinetic energy with standard
deviations.}\label{fig:e4}
\end{figure}

The~control over the flow that is available with two-band forcing is
examined further by investigating the~effects of varying
the~strength of the forcing in the second band. We~kept the~energy
input rate for the~first $k \leq 3\pi$ band equal to $\varepsilon_w
= 0.15$ and varied the~strength of forcing in the~second $33\pi < k
\leq 41\pi$ band adopting $\varepsilon_w = 0.075 \ldots 0.90$.
The~corresponding compensated energy spectra from these simulations
are shown in Fig.~\ref{fig:e4}(a). We~observe that a~higher energy
input-rate in the second band leads to a more pronounced peak in the
spectrum which shifts to lower values of $k\eta$ with increasing
$\varepsilon_{w}$ of the second band. Simultaneously, the value of
${E}(k)$ decreases for the low-$k$ modes with increasing
$\varepsilon_{w}$. This forcing of the~second band allows to quite
independently control the spectrum, at roughly the same total energy
content in the~flow. In~fact, variation of $\varepsilon_{w}$ of the
second band by a~factor of about 10 is seen to lead to a~comparably
strong increase in the~peak value of the spectrum while the~total
energy level $\langle \widehat{E} \rangle_{t}$ is increased by only
$\approx 15 \%$ as seen in Fig.~\ref{fig:e4}(b).

In the~next section we will examine how the~changes in the flow
properties due to the two-band forcing in {\it{spectral space}} influence
the~{\it{physical space}} mixing efficiency of a~passive scalar.

\section{Small and large scale mixing efficiency}
\label{mixeff}

The~consequences of explicit broad band forcing not only express
themselves in modulated energy cascades. The mixing properties of
the evolving turbulent flow in physical space also depend
significantly on the~forcing that is applied. In this section we
quantify the~mixing-efficiency by monitoring geometric properties of
evolving level-sets of an~embedded passive scalar. The numerical
integration method that is used to determine these level-set
properties is described in subsection \ref{levelset}.
The~ensemble-averaged simulation results are discussed in subsection
\ref{surfwrink}; we establish to what extent the two-band forcing
can be used to control the maximal rate of mixing and the total
accumulated degree of mixing.

\subsection{Level-set evaluation to quantify mixing}
\label{levelset}

To~illustrate and quantify the~influence of two-band forcing on
the~turbulent dispersion of a~passive scalar field we~analyze the
evolution of basic geometric properties of its level-sets.
As~a~result of the turbulent flow these level-sets become highly
distorted and dispersed across the flow-domain. Specifically,
we~concentrate on the~surface-area and the~wrinkling of these
level-sets. We~adopt a~specialized integration method to determine
these geometric properties, as developed in
\cite{geurts:mixing:2001}. This method is based on the Laplace
transform and avoids the explicit construction and integration over
the~complex and possibly fragmented scalar level-sets. With this
method an~accurate and efficient evaluation of the~evolving mixing
efficiency can be achieved which allows to quantify the~increased
complexity of the~flow in relation to the~two-band forcing that is
used.

Basic geometric properties of a level-set $S(a,t)= \{ \mathbf{x} \in
\mathbb{R}~|~C(\mathbf{x},t)=a \}$ of the scalar $C({\mathbf{x}},t)$
may be evaluated by integrating a corresponding `density function'
$g$ over this set. In fact, we have:
\begin{equation}
I_{g}(a,t) = \int_{S(a,t)} dA~g(\mathbf{x},t)
           = \int_{V} d\mathbf{x} ~\delta(C(\mathbf{x},t)-a) | \nabla C(\mathbf{x},t)|g(\mathbf{x},t)
\end{equation}
where the volume $V$ encloses the level-set
$S(a,t)$~\cite{mazja:1985}. Setting \mbox{$g(\mathbf{x},t) = 1$},
\mbox{$g(\mathbf{x},t)=\nabla \cdot {\mathbf{n}(\mathbf{x},t)}$} or
$g(\mathbf{x},t)=\left| {\nabla \cdot {\mathbf{n}(\mathbf{x,t})}}
\right|$, we~can determine the~global surface-area, curvature or
wrinkling of $S(a,t)$ respectively. Here
\mbox{$\mathbf{n}(\mathbf{x},t) =\nabla C(\mathbf{x},t) / |\nabla
C(\mathbf{x},t)|$} is a~unit normal vector, locally perpendicular to
the~level-set. The~divergence of this vector-field is directly
related to the~local curvature of the~level-set.

We~will focus on the~evolution of the~surface-area $A$ and
the~wrinkling $W$. The scalar $C$ is scaled to be between 0 and 1;
we~will primarily consider the~level-set $a=1/4$. In particular
we~monitor:
\begin{equation}\label{eq:aw}
\vartheta_{A}(a,t) = \frac{I_A(a,t)}{I_A(a,0)} ~~~;~~~ \vartheta_{W}(a,t) = \frac{I_W(a,t)}{I_W(a,0)}
\end{equation}
By determining $\vartheta_{A}$ and $\vartheta_{W}$ we may quantify
the rate at which surface-area and wrinkling develop, the maximal
values that are obtained and the~time-scale at which these are
achieved. The~corresponding cumulative effects are given by
\begin{equation}\label{eq:caw}
\zeta_{A}(a,t) = \int_0^t {\vartheta_{A}(a,\tau) d\tau} ~~~;~~~ \zeta_{W}(a,t) = \int_0^t {\vartheta_{W}(a,\tau) d\tau}
\end{equation}
These cumulative measures express the total surface-area and
wrinkling that has developed in the course of time. In particular
applications, e.g., involving combustion in diffusion flames, the
cumulative surface-area and wrinkling express the total `chemical
processing capacity'. Here, we will determine these cumulative
effects in order to characterize the different two-band forcing
procedures.

To establish the influence of forcing on turbulent mixing properties
we~simulated the~spreading of a~passive tracer at Schmidt number
$Sc=0.7$. The simulations were started from a~spherical tracer
distribution of radius $r=3/16$. The scalar concentration was set
equal to 1 inside this sphere and 0 outside. {A~localized
Gaussian smoothing of this $C$-distribution was applied} near the
edge of the initial sphere to avoid resolution problems. The~fractal
forcing B2-procedure  as defined in (\ref{eq:mazzi}) was adopted.
We~performed numerical simulations in which the~energy input-rate
$\varepsilon_{w}$ and the~spectral support of this two-band forcing
were varied.

\begin{figure}
\begin{center}
\includegraphics[width=0.29\textwidth]{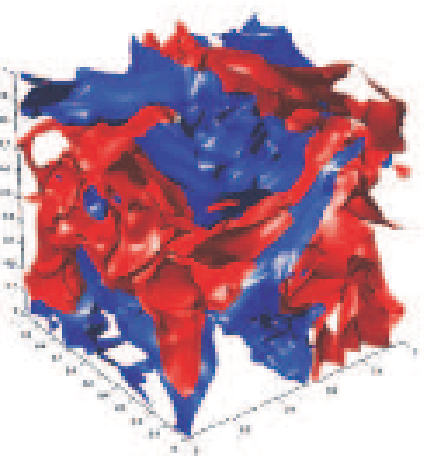}
\includegraphics[width=0.29\textwidth]{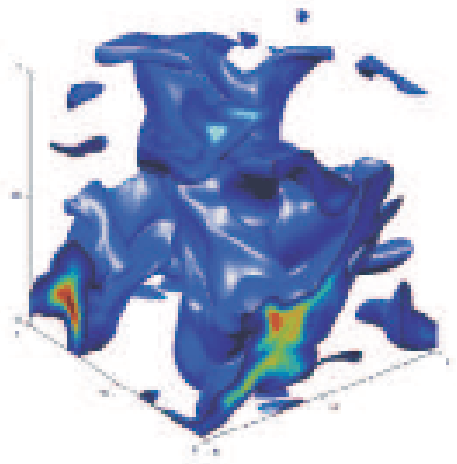}(a)\\
\includegraphics[width=0.29\textwidth]{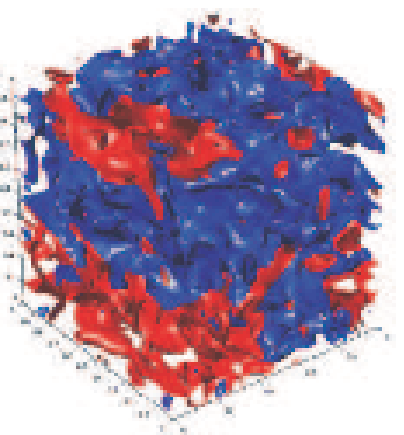}
\includegraphics[width=0.29\textwidth]{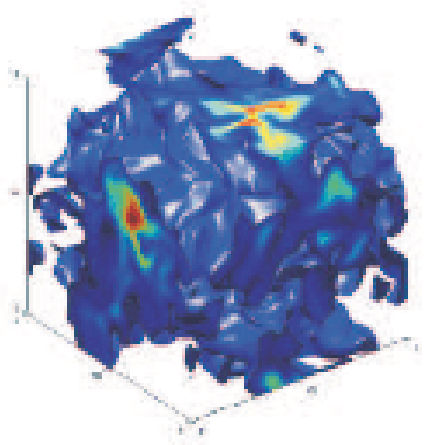}(b)\\
\includegraphics[width=0.29\textwidth]{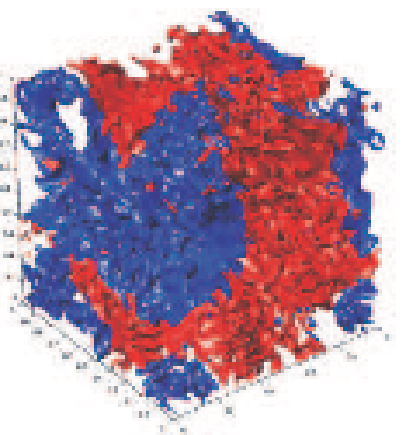}
\includegraphics[width=0.29\textwidth]{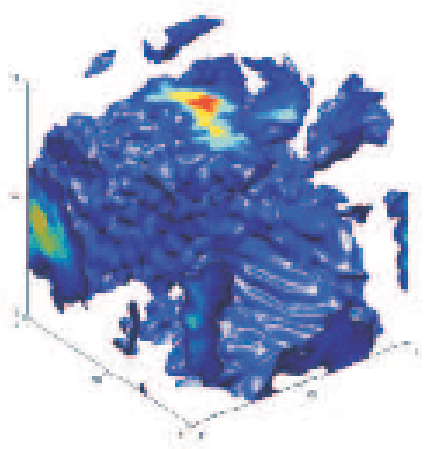}(c)
\end{center}
\caption{Snapshot of vertical velocity field iso-surfaces (left) and
passive scalar concentration (right) at $t=0.5$ for large-scale
forcing $\mathbb{K}_{1,1}$ with $\varepsilon_w=0.6$ (a), or with
$\varepsilon_{w,1}=0.15$ in the first shell and complementary
forcing $\varepsilon_{w,2}=0.45$ in $\mathbb{K}_{5,8}$ (b) or
$\mathbb{K}_{13,16}$ (c). In the velocity field snapshots the red
iso-surface corresponds to $u_{2}=0.2$ and the blue iso-surfaces to
$u_{2}=-0.2$. The iso-surface at $C=0.25$ is shown for the passive
scalar.} \label{fig:snap}
\end{figure}

As point of reference we~adopted large-scale forcing in
the~${\mathbb{K}}_{1,1}$ shell with an~energy injection-rate
$\varepsilon_w=0.6$.  The~resolution requirements were
satisfactorily fulfilled: $k_{\max}\eta$ ranges from $2.3$ to $3.5$
using a~resolution in the range $128^3 - 192^3$ grid-cells.
For~the~passive scalar these resolutions correspond to
$k_{\max}\eta_{OC}$ in the range from $3$ to $4.5 $, where
$\eta_{OC}$ is the~Obukhov-Corrsin scale
\cite{watanabe:statistics:2004}. To~study the influence of two-band
B2-forcing we applied supplementary forcing either in a~region
situated near the~largest scales of~the~flow, i.e.,
${\mathbb{K}}_{5,8}$ or further separated, i.e.,
${\mathbb{K}}_{13,16}$. In case two bands are forced, the~energy
input-rate for the $\mathbb{K}_{1,1}$ shell is
$\varepsilon_{w,1}=0.15$, while the second band is forced using
$\varepsilon_{w,2}=0.45$. In this way the total energy level is kept
at comparable levels in the different cases. A~qualitative
impression of the effect of these forcing procedures may be observed
from the snapshots shown in Fig.~\ref{fig:snap}. The~velocity and
passive scalar display considerably more small-scale features in
case of two-band forcing, particularly in case of high-$k$ forcing.
To quantify this qualitative impression we~apply the~level-set
analysis discussed above. The~results will be presented in the~next
subsection.

\subsection{Surface-area and wrinkling}
\label{surfwrink}

In this subsection, we compare instantaneous and accumulated mixing
properties for large-scale forcing and different two-band forcing.
The total energy input rate to the flow is kept constant at $0.6$;
a~fraction $\varepsilon_{w,1}$ is allocated to the first shell and
$\varepsilon_{w,2}$ to the second band such that
$\varepsilon_{w,1}+\varepsilon_{w,2}=0.6$ and $\varepsilon_{w,1}$ is
varied from $0.05$ up to $0.6$. The characterization of the
mixing-efficiency was based on averaging $20$ simulations, each
starting from an independent realization of the~initial velocity
field. The~different initial conditions were each separated by two
eddy-turnover times.

\begin{figure}
\begin{center}
\includegraphics[width=0.45\textwidth]{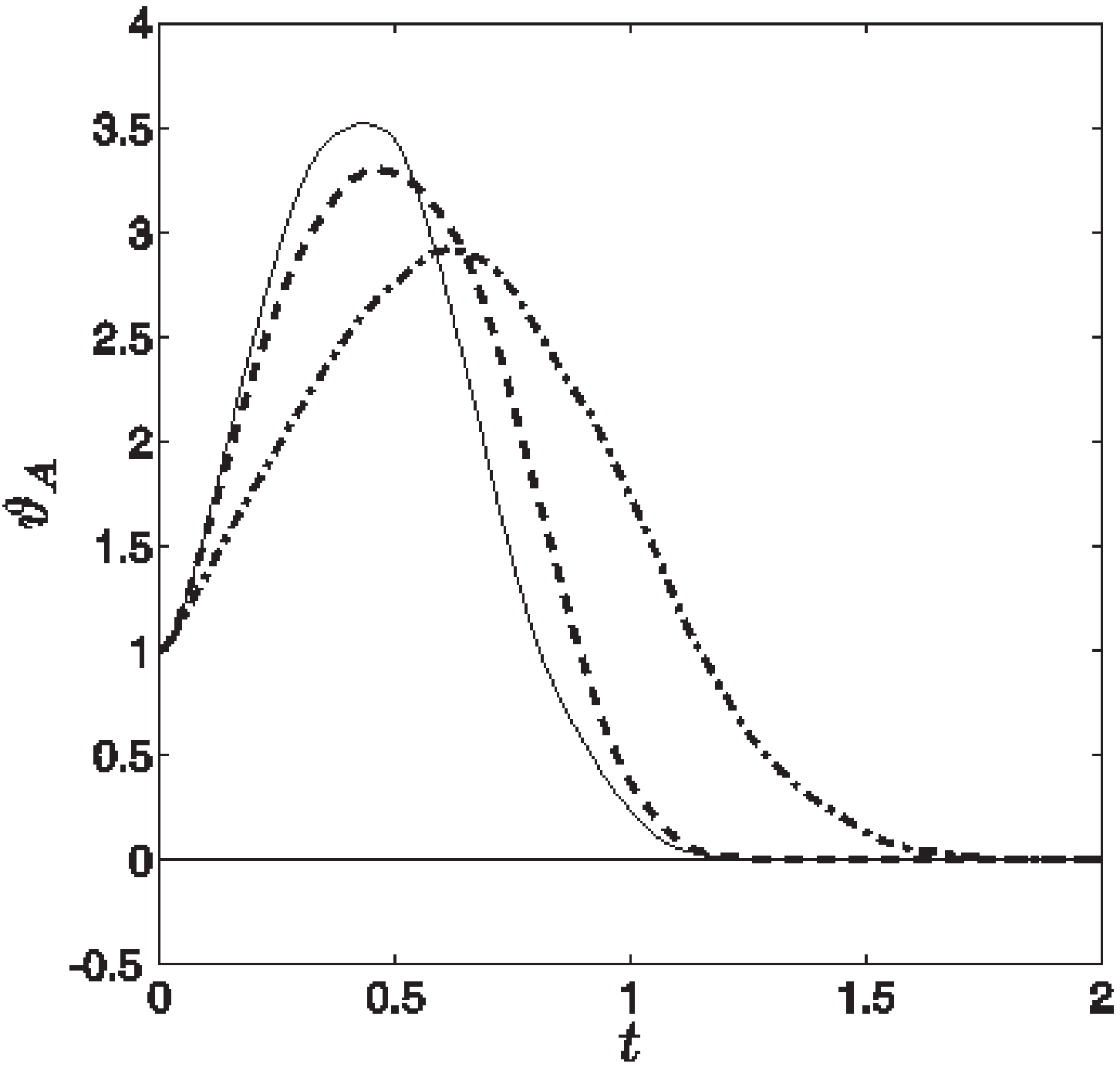}(a)
\includegraphics[width=0.44\textwidth]{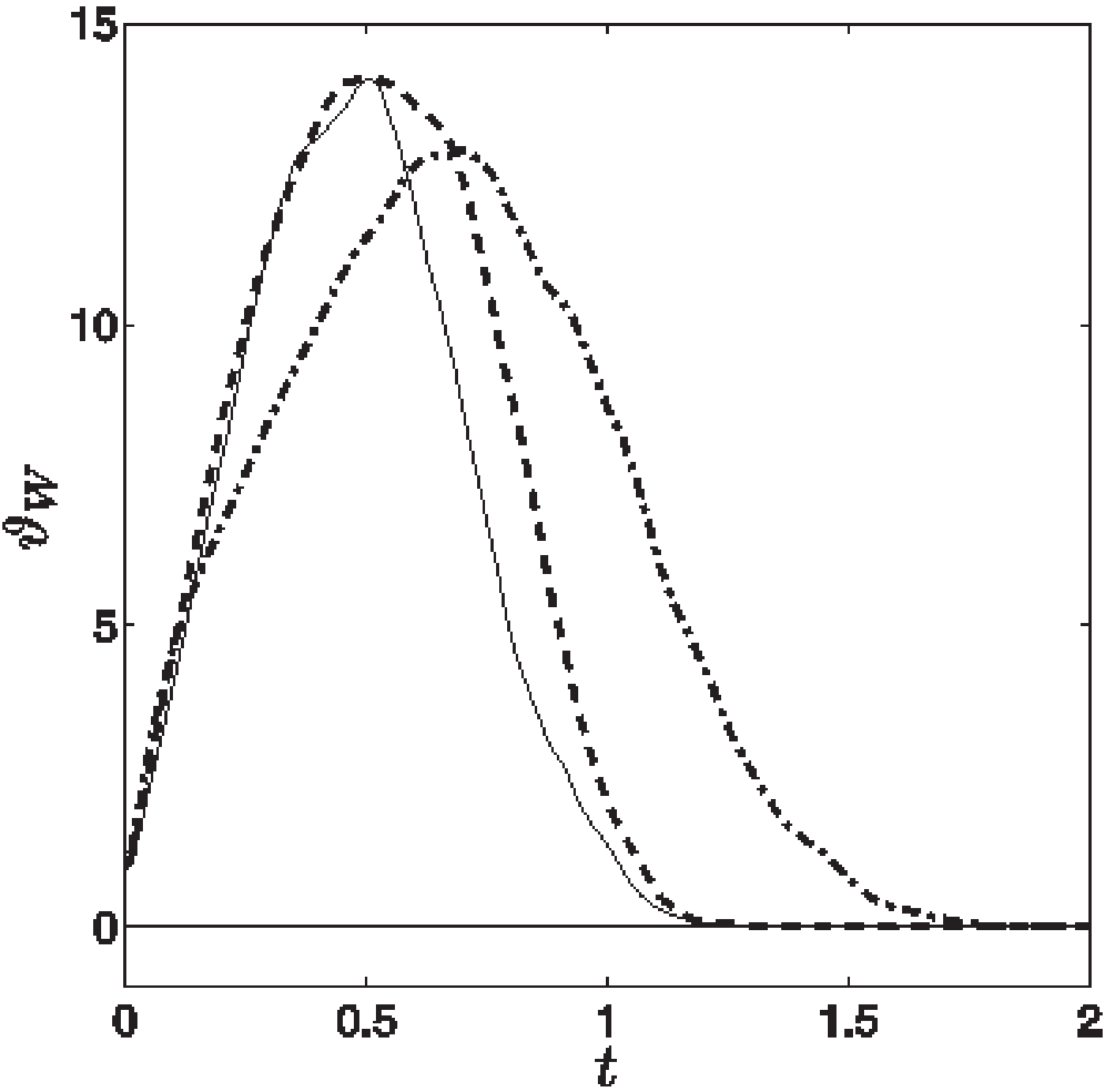}(b)\\
\includegraphics[width=0.45\textwidth]{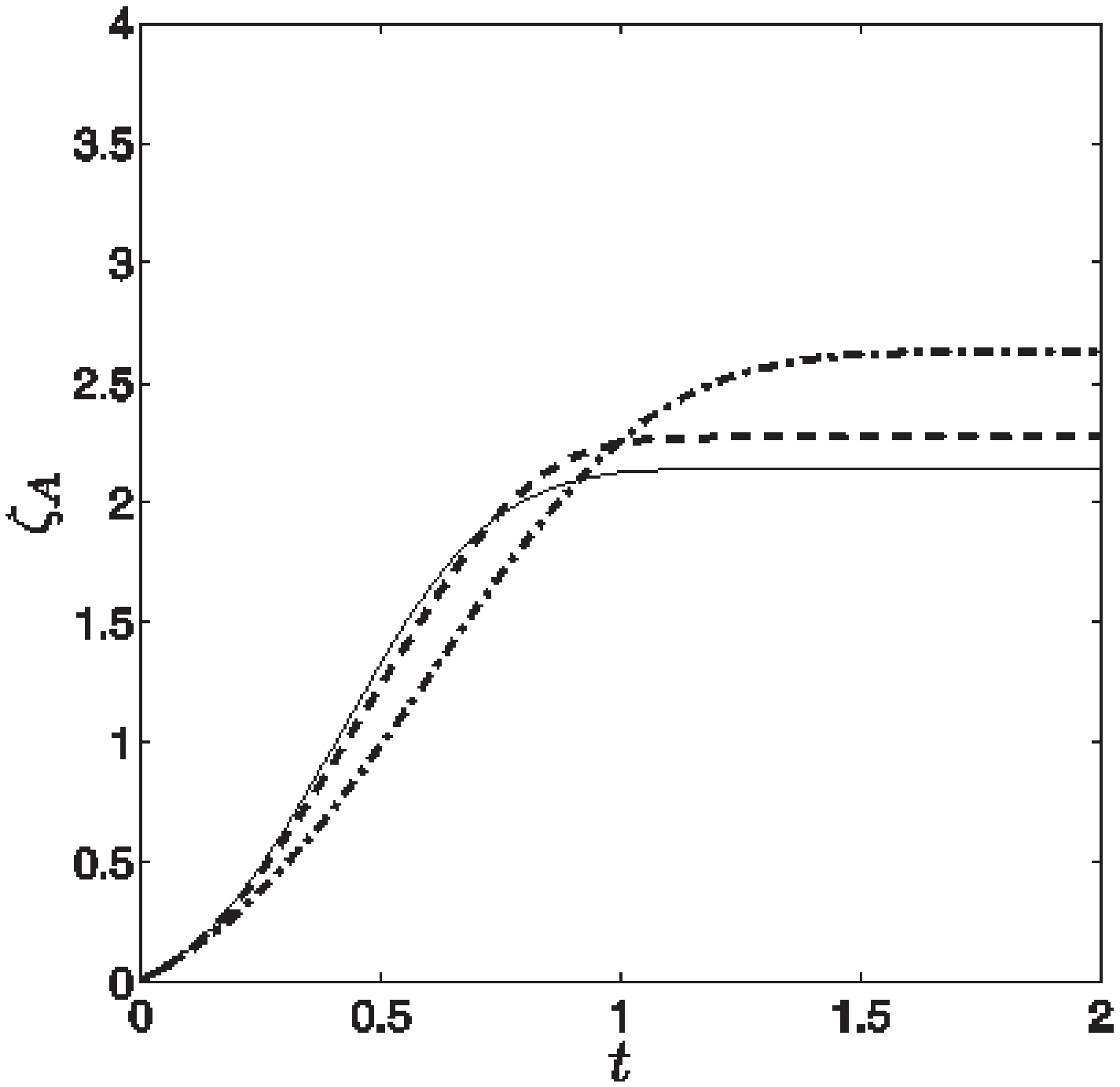}(c)
\includegraphics[width=0.44\textwidth]{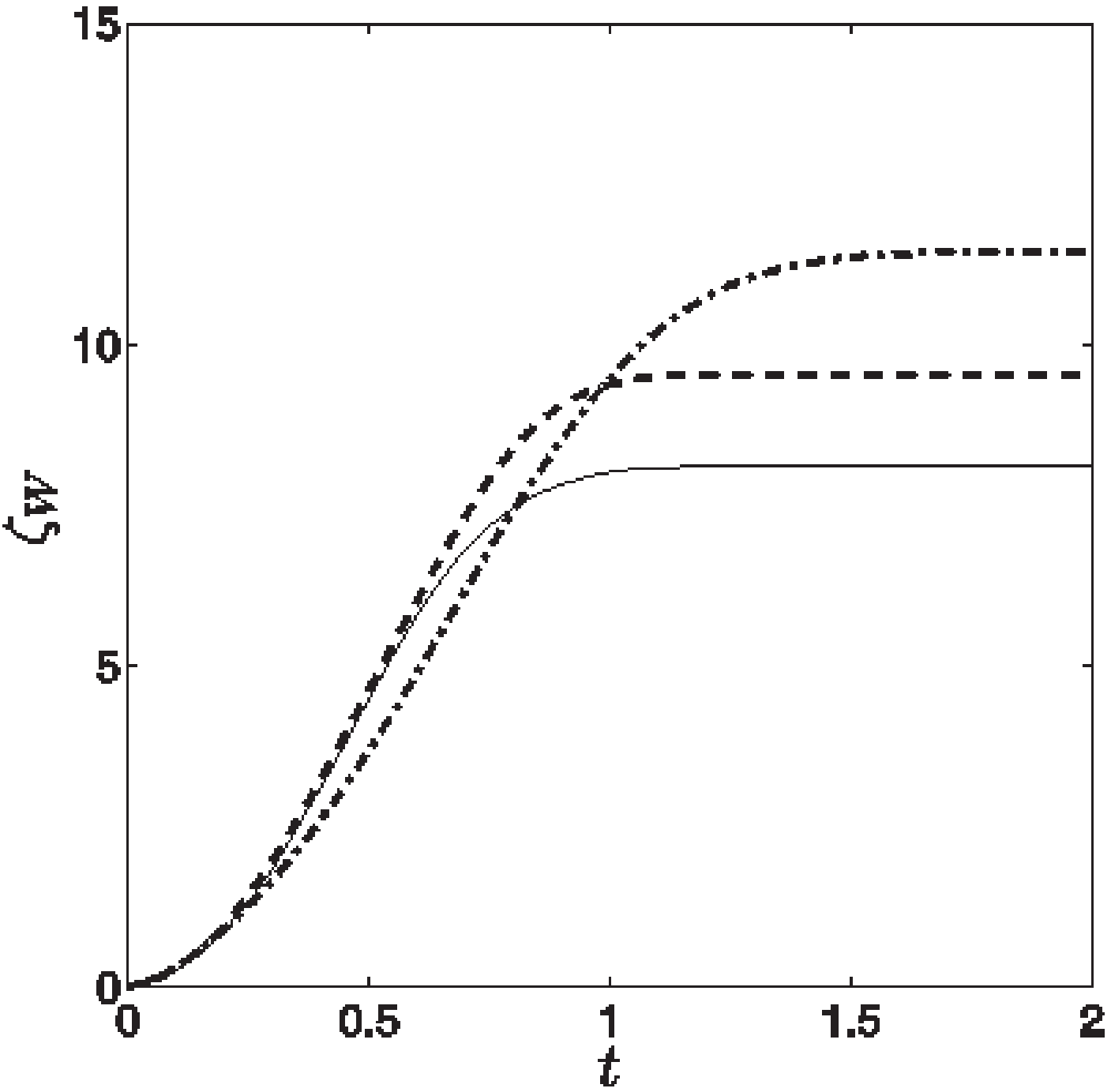}(d)
\end{center}
\caption{Evolution of decaying passive scalar growth parameters:
a)~surface-area $\vartheta_A$, b)~wrinkling $\vartheta_W$,
c)~accumulated surface-area $\zeta_A$, d)~accumulated wrinkling
$\zeta_W$. Large-scale forcing $\mathbb{K}_{1,1}$ with
$\varepsilon_{w}=0.60$~(solid) and complementary two-band forcing
($\varepsilon_{w,1}=0.15$ and $\varepsilon_{w,2}=0.45$) in
$\mathbb{K}_{5,8}$~(dashed),
$\mathbb{K}_{13,16}$~(dash-dotted).}\label{fig:area1}
\end{figure}

The~instantaneous and cumulative effects arising from both the
large-scale and the two-band forcing are shown in
Fig.~\ref{fig:area1}. The~development of the instantaneous
surface-area and wrinkling is qualitatively similar in each case.
The~concentrated initial tracer distribution is in the~first stages
primarily dispersed by convective sweeping in the turbulent flow. As
a~result, the~level-set corresponding to $a=1/4$ becomes distorted
and both $\vartheta_{A}$ and $\vartheta_{W}$ show a rapid increase.
However, since no source of scalar was included in the computational
model, molecular diffusion dominates the long-time behavior and
leads to $\vartheta_{A}$ and $\vartheta_{W}$ to decrease to zero
as~~$t \rightarrow \infty$. In~between, $\vartheta_{A}$ and
$\vartheta_{W}$ reach their maximum. The~rapid initial growth is
also clearly expressed in Fig.~\ref{fig:area1}(c-d). In~addition,
the~cumulative measures $\zeta_{A}$ and $\zeta_{W}$ show a clear
saturation as $t \gtrsim 1$.

We~observe from Fig.~\ref{fig:area1}(a-b) that forcing of the large
scales only, creates the highest growth-rate of surface-area and
wrinkling. The surface-area reaches its maximum value both sooner
and at a~higher value in this case. In the~initial stages convective
spreading of the~tracer dominates over the~decay caused by molecular
diffusion; hence in these stages the~agitation of the~larger scales plays a~crucial
role in the evolution of the surface-area. The~higher band forcing
needs to compete more directly with the viscous effects and does not
induce very strong sweeping motions over large distances.
Correspondingly, high-$k$ forcing is found less effective in
producing surface-area. The more localized distortions of the~scalar
level-sets, as expressed by the development of the~wrinkling, are
less affected by the~competition with viscosity, as seen in
Fig.~\ref{fig:area1}(b).

The interpretation of the effectiveness of the mixing in relation to
the specific forcing alters if we compare the accumulated values for
surface-area and wrinkling. As may be appreciated from
Fig.~\ref{fig:area1}(c-d), a significant enhancement of the
accumulated long-time surface-area and wrinkling arises as a~result
of the explicit agitation of the smaller scales in the flow.
Evidently, forcing of the smaller scales does not yield a more
intense mixing, judging from the instantaneous values, but does
yield an increase in the total surface-area and wrinkling,
accumulated over time.

\begin{figure}
\begin{center}
\includegraphics[width=0.45\textwidth]{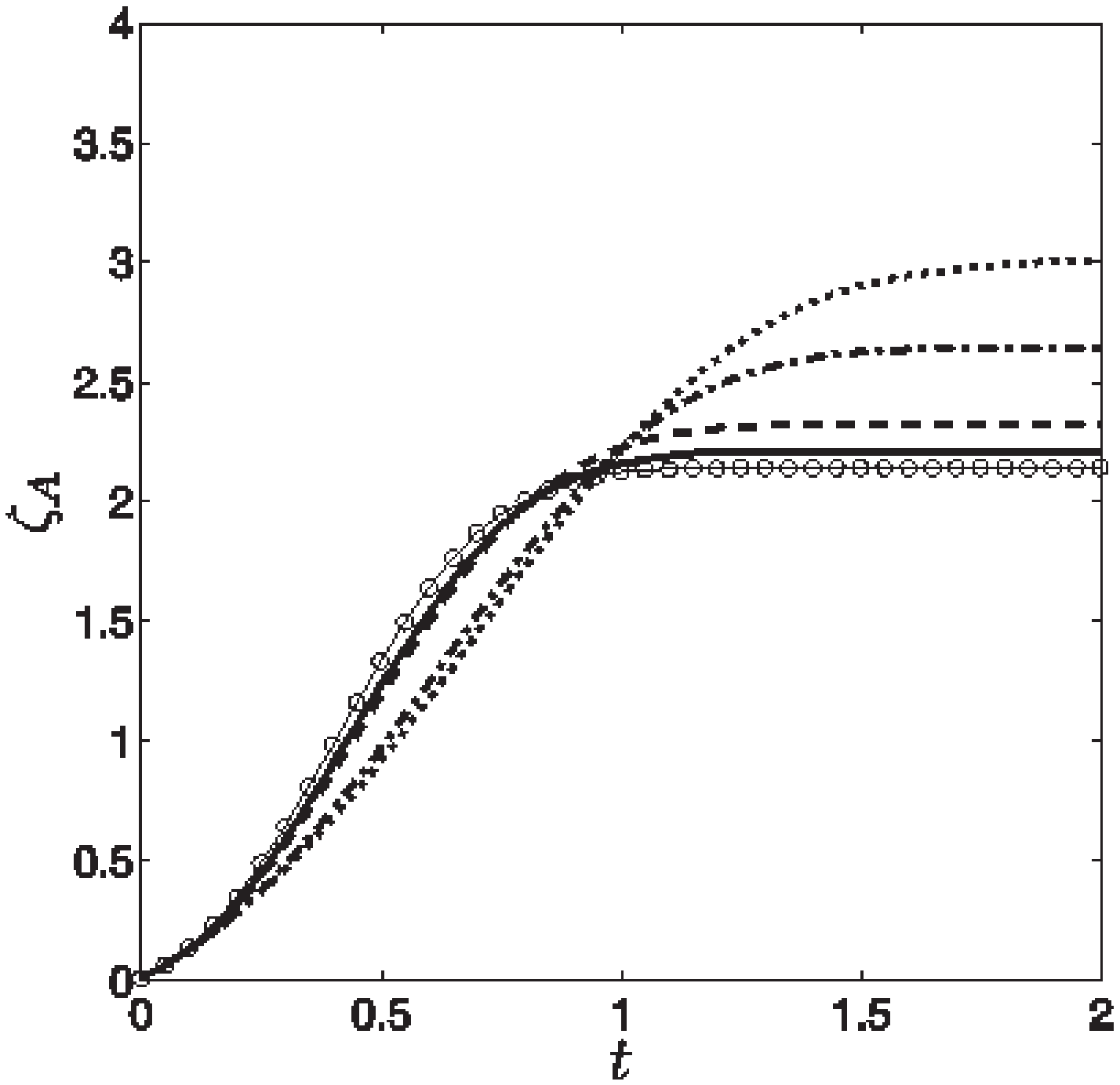}(a)
\includegraphics[width=0.44\textwidth]{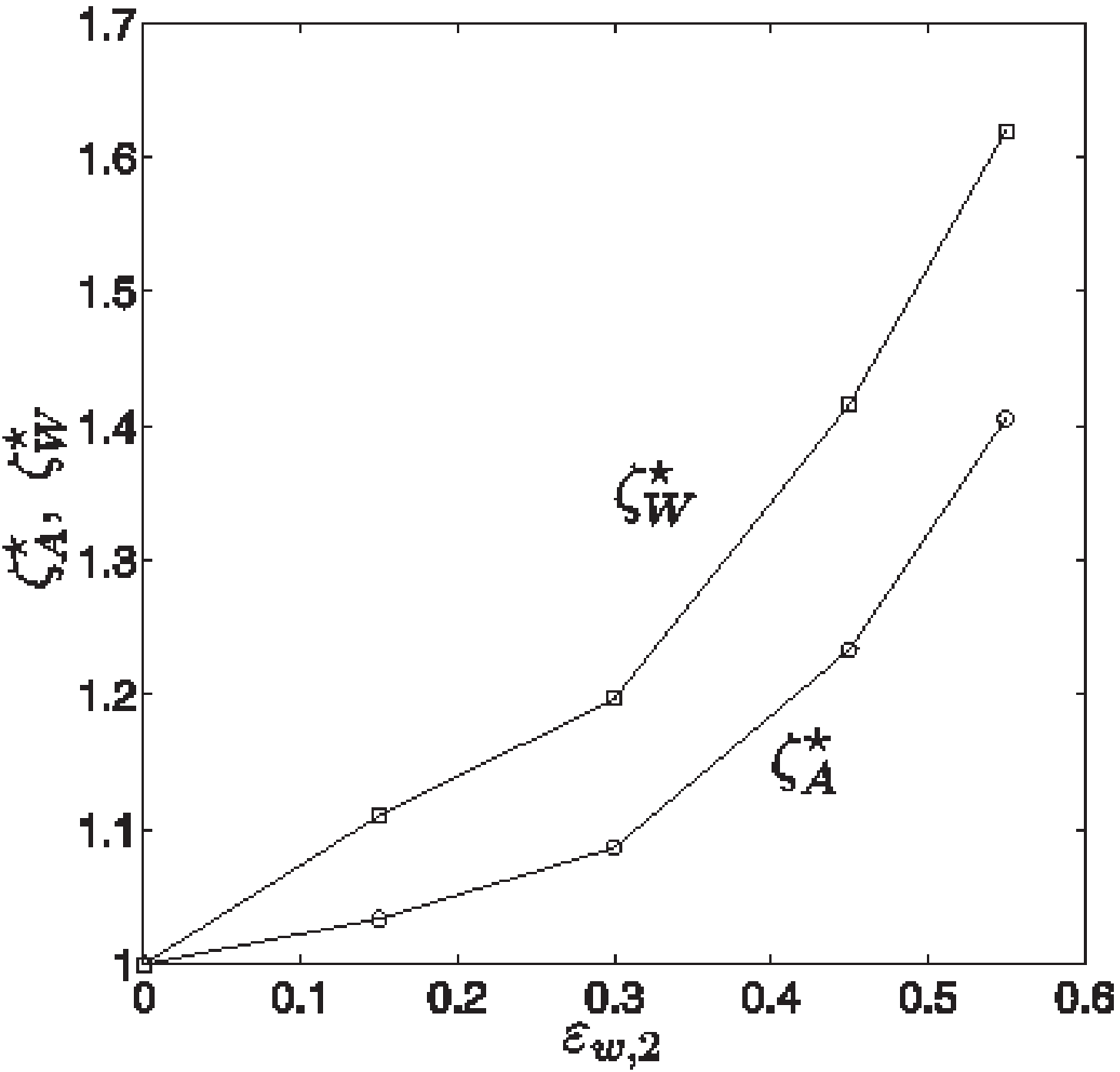}(b)
\end{center}
\caption{(a) Evolution of the decaying passive scalar accumulated
surface-area parameter~$\zeta_A$ for~two-band forcing
($\mathbb{K}_{1,1}$ -- $\mathbb{K}_{13,16}$) with different
equi-partitions of energy between bands:
$(0.60$~--~$0.00)$~($\circ$), $(0.45$~--~$0.15)$~(solid),
$(0.30$~--~$0.30)$~(dashed), $(0.15$~--~$0.45)$~(dot-dashed),
$(0.05$~--~$0.55)$~(dotted). (b) Total
surface-area~$\zeta^{\star}_A$ and wrinkling~$\zeta^{\star}_W$ at
$t=2$ for different $\varepsilon_{w,2}$ in $\mathbb{K}_{13,16}$
(results normalized by the total surface-area and wrinkling for
the~large-scale forcing).} \label{fig:area2}
\end{figure}

To measure the~influence of variations in the strength of
the~forcing in the~high-$k$ band, we~focus on the~$\mathbb{K}_{1,1}$
and $\mathbb{K}_{13,16}$ two-band forcing. In particular, we~keep
$\varepsilon_{w,1}+\varepsilon_{w,2}=0.6$ and vary the values of
$\varepsilon_{w,2}$. The effects on the cumulative mixing-efficiency
are shown in Fig.~\ref{fig:area2}(a). We observe that an increase in
$\varepsilon_{w,2}$ implies a slight decrease in the initial growth
rate of instantaneous surface-area, but an increase in the long-time
cumulative effect. The dependence of the long-time cumulative effect
on $\varepsilon_{w,2}$ is clarified in Fig.~\ref{fig:area2}(b).
These simulation results establish the degree of control that may be
achieved with two-band forcing and the feasibility of such
computational modeling. This approach may help to identify optimal
stirring procedures to which future research will be devoted.

\section{Summary and concluding remarks} \label{concl}

Various deterministic forcing methods that perturb a~turbulent flow
in a chosen range of length-scales were examined. The presented
modeling framework incorporates the~explicit forcing as an integral
part. We~have shown that with a~relatively simple forcing model
basic properties of complex flows can be captured. For example, an
enhancement of the energy dissipation by small-scale forcing was
seen to produce so-called spectral short-cut features, quite similar
to what was observed experimentally in flow over canopies
\cite{finnigan:turbulence:2000} where the~kinetic energy is
immediately transferred to the~smallest scales of the~flow.

Forcing methods agitating the~flow in a wide range of scales induce
significant differences in the developing flow, compared to the case
obtained classically in which only the~large scales are forced.
Various forcing methods were introduced and shown to produce
qualitatively similar results, provided the~forcing parameters
correspond to turbulence at comparable total kinetic energies. We
classified the methods according to constant-energy or
constant-energy-input-rate and examined these procedures by
simulating forced turbulence with energy injected in two different
bands. The modulation of the turbulent flow was investigated for
various locations of the second high-$k$ band in spectral space. It
was shown that the forcing in the second band induces nonlocal
modulation of the energy spectrum. This was further examined by
simulations done with different strength of forcing in the high-$k$
band controlled by the energy injection rate.

We devoted special attention to a~recently proposed multiscale
forcing that models a~flow under the influence of an additional
perturbation by a~multiscale object \cite{mazzi:fractal:2004}.
We~performed numerical simulations of the dispersion of a passive
scalar field in a turbulent flow that is driven by such forcing.
A~level-set integration method was adopted to quantify general
characteristics of mixing quality and efficiency. It~was found that
broad-band forcing causes additional production of smaller scales in
the~flow which is directly responsible for the~localized enhancement
of the wrinkling of the level-set. In contrast, the surface-area of
a~level-set of the tracer is found to be mainly governed by
convective sweeping by the~larger scales in the flow and hence it is
governed to a greater extent by the energy input-rate allocated to
the small-$k$ range. Future study will include the spatial
localization of the forcing. This can help to model flows that are
more closely related to realistic physical situation observed in
experiments and applications.

\section*{Acknowledgments}
The~work is supported by the Foundation for Fundamental Research on
Matter (FOM), Utrecht, the Netherlands. Simulations were performed
at SARA; special thanks go to Willem Vermin for support with the
parallelization. The~computations were made possible through grant
SC-213 of the~Dutch National Computing Foundation (NCF). AKK~would
like to thank David McComb for many stimulating discussions during
research visits at the~University of Edinburgh. BJG~gratefully
acknowledges support from the~Turbulence Working Group (TWG) at the
Center for Non-Linear Studies (CNLS) at Los Alamos National
Laboratory which facilitated an extended research visit in 2005 and
allowed many fruitful discussions. Special thanks go to Darryl Holm.

\appendix
\section{}

To eradicate the~aliasing error we~study in more detail {\it{(i)}}
the random phase shifts method and {\it{(ii)}} the method employing
two-shifted grids with spherical truncation
\cite{rogallo:illiac:1977,rogallo:numerical:1981,canuto:spectral:1988}.
In the~first case, the~aliasing error is only partially removed.
With additional truncation of the~Fourier velocity field
coefficients the~remaining error can be reduced to
$\mathcal{O}(\Delta t^2)$. In the method employing two-shifted grids
and spherical truncation, the~aliasing error can be fully removed
from the~simulations. This specific approach doubles the
computational costs and memory requirements, compared to the~random
shifts method. The~well-known $3/2$ method can be used as well, by
going to higher resolution and truncating the field. This can be
done with the~lowest number of operations, but has higher memory
requirements.

\begin{figure}[hbt]
\centering{
\includegraphics[width=0.45\textwidth]{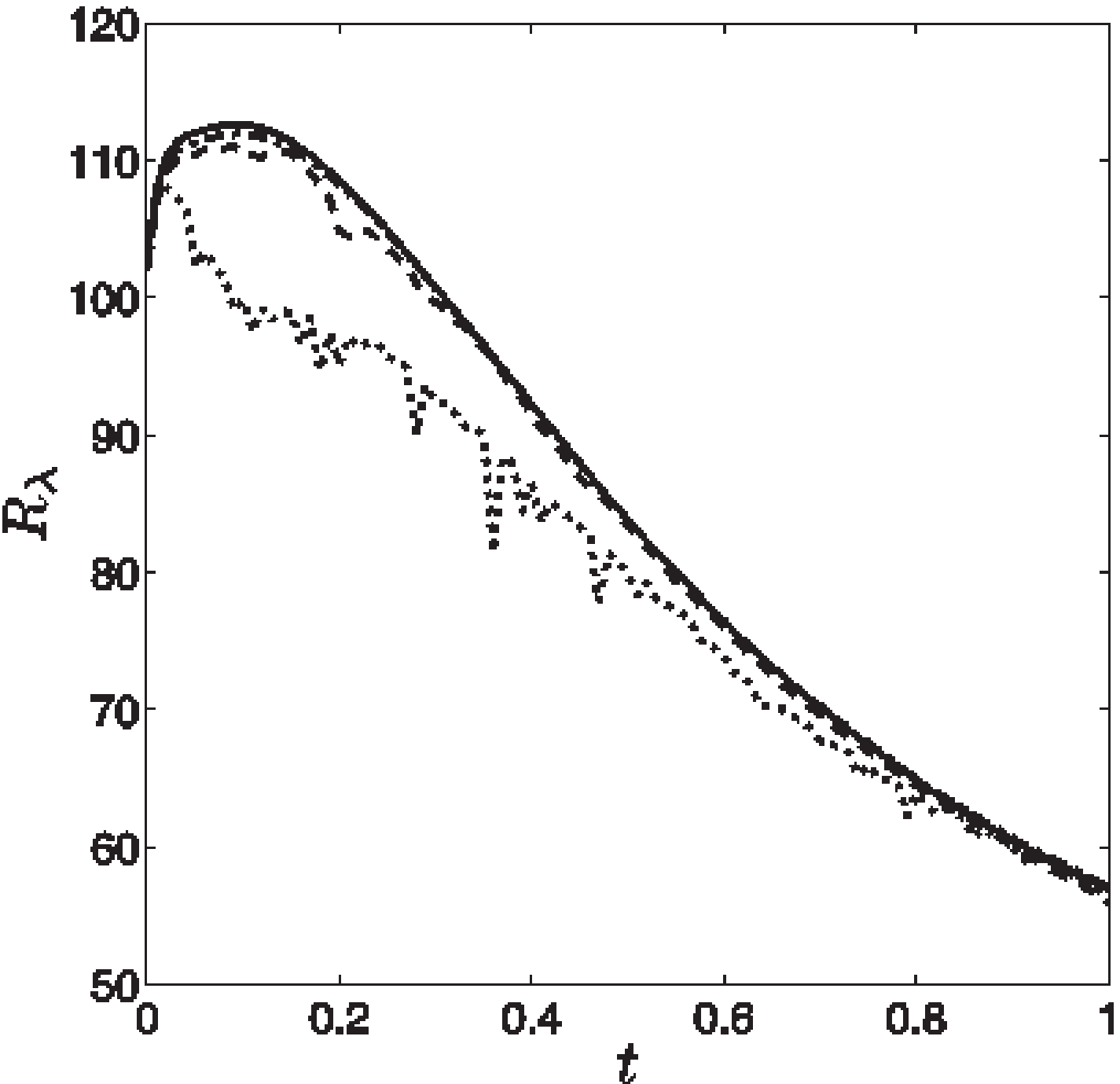}(a)
\includegraphics[width=0.45\textwidth]{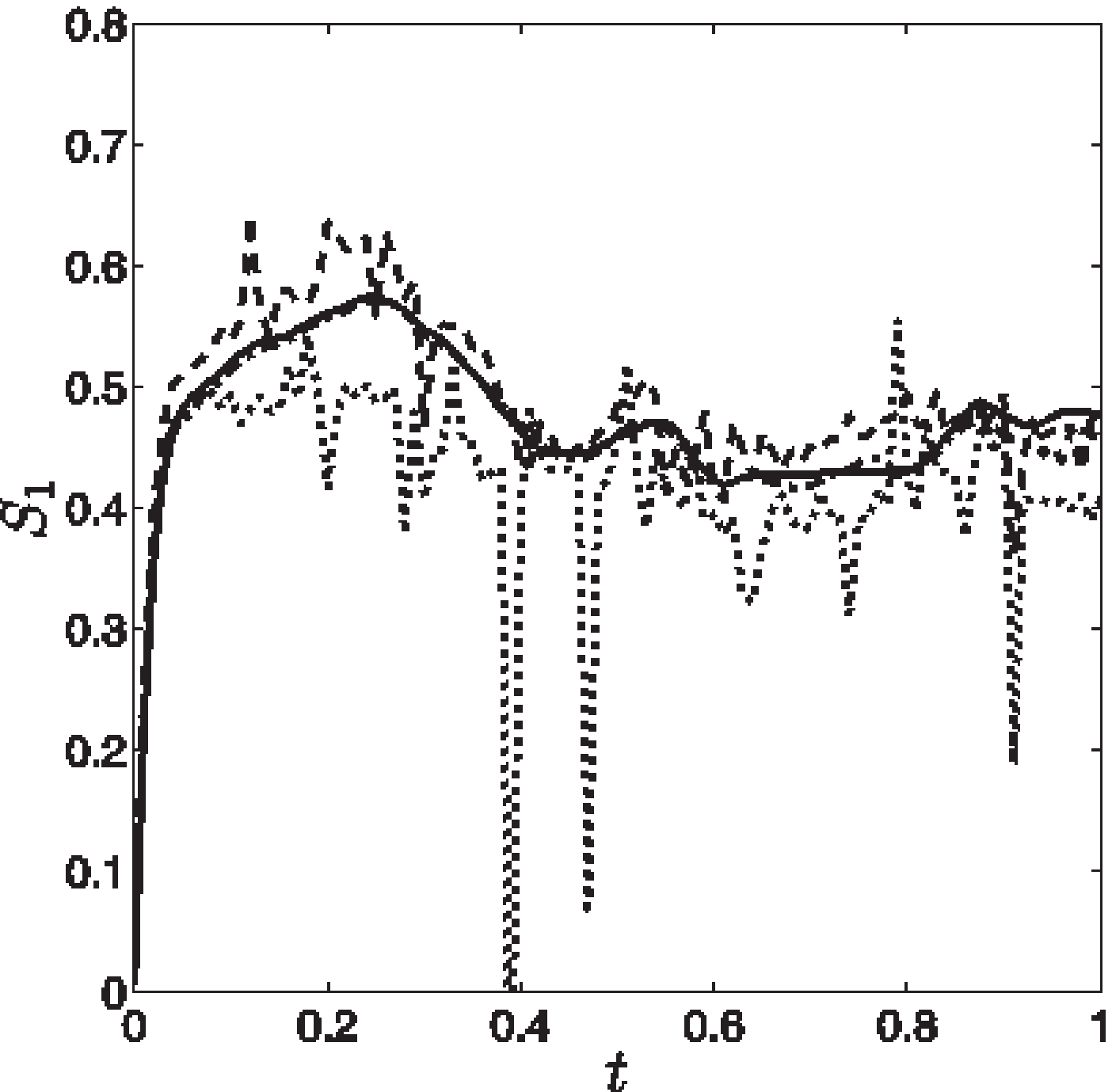}(b)\\
\includegraphics[width=0.45\textwidth]{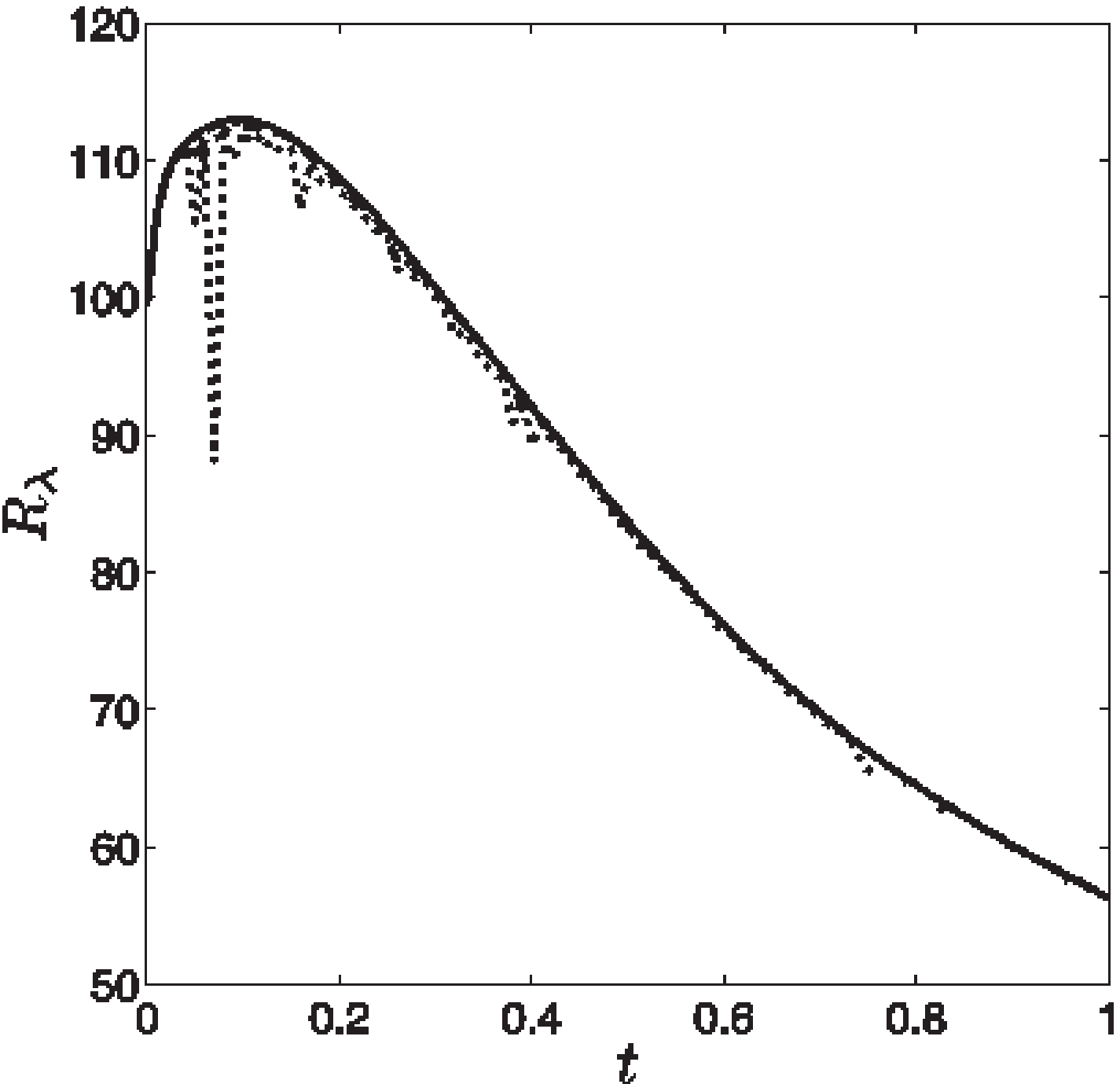}(c)
\includegraphics[width=0.45\textwidth]{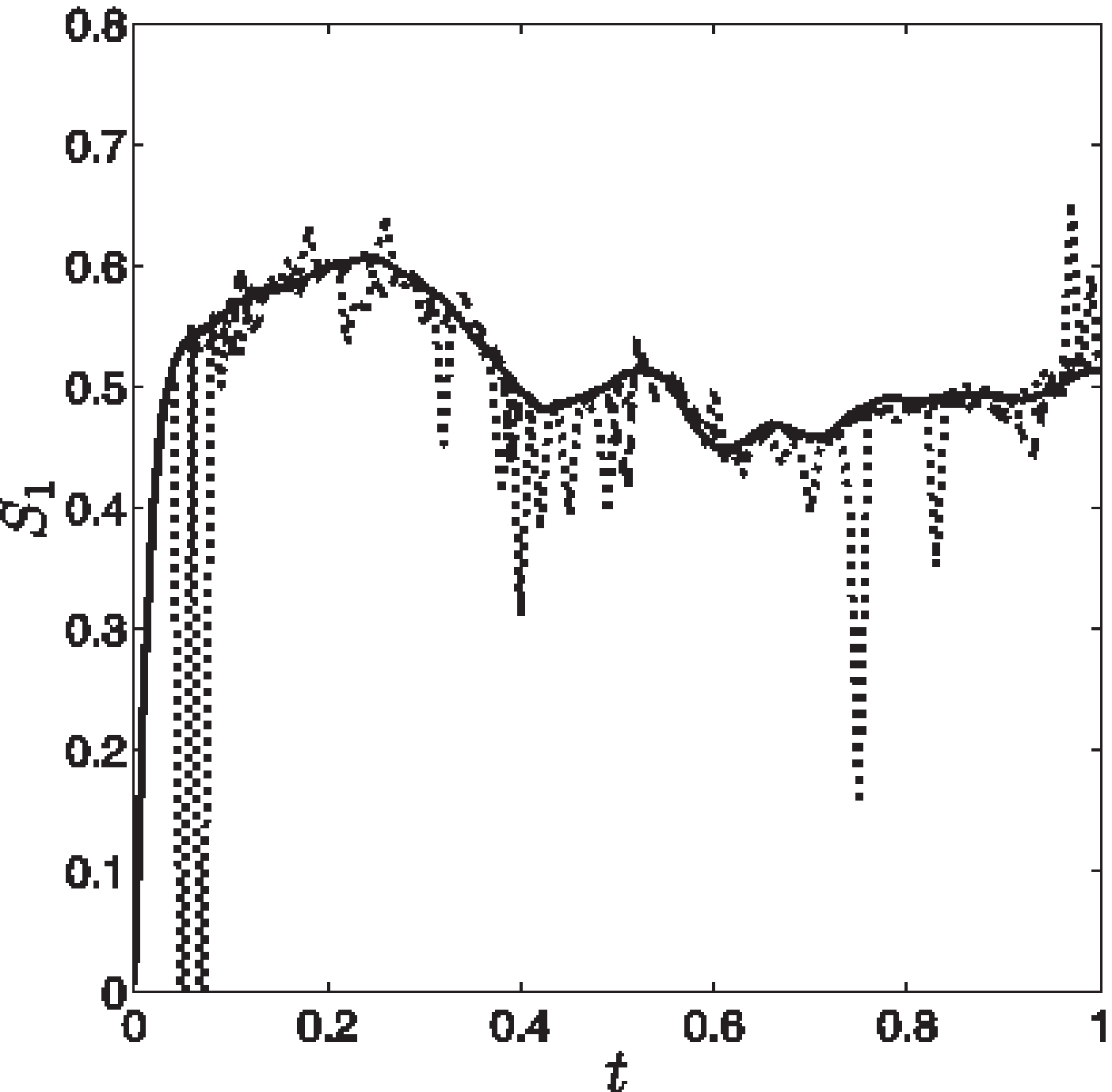}(d)}
\caption{Influence of aliasing error for resolution $128^3$ (a-b)
and $192^3$~(c-d) on Taylor-Reynolds number $R_\lambda(t)$
and~longitudinal skewness $S_1(t)$ at an initial $R_{\lambda}=100$
case. Simulations with aliasing error (dotted), partial dealiasing
without truncation (dashed), partial dealiasing with the~truncation
(dash-dotted), full dealiasing by two grid shifts (solid). Results
for the~partial dealiasing with truncation (dash-dotted) are almost
identical to fully dealiased results (solid).} \label{fig:aliasing}
\end{figure}
The~aliasing error for higher-resolution runs affects mainly
the~small-scale statistics. This is visualized
in~Fig.~\ref{fig:aliasing} where we~have shown the~Taylor-Reynolds
number and the longitudinal skewness for decaying turbulence simulations
with initial $R_\lambda=100$ and two resolutions $128^3$ and
$192^3$. The partial
dealiasing removes the~main aliasing error and with the~additional
truncation reduces it to the~accuracy associated with the adopted
Runge-Kutta scheme. There is a~small difference between
the~full and partial removal of the aliasing error for the~lower resolution of $128^3$,
but this largely vanishes for the~well-resolved $192^3$ case.

\bibliographystyle{plain}
\bibliography{mix-man}

\begin{thebibliography}{10}

\bibitem{banhart:cellular:2001}
J.~Banhart, M.F. Ashby, and N.A. Fleck.
\newblock {\em Cellular metals and metal foaming technology}.
\newblock Verlag MIT, Bremen, 2001.

\bibitem{batchelor:theory:1953}
G.~K. Batchelor.
\newblock {\em Theory of homogeneous turbulence}.
\newblock Cambridge University Press, 1953.

\bibitem{boomsma:metal:2002}
K.~Boomsma, D.~Poulikakos, and F.~Zwick.
\newblock Metal foams as compact high performance heat exchangers.
\newblock {\em Mech. Mater.}, 35:1161--1176, 2003.

\bibitem{breugem:direct:2005}
W.~P. Breugem and B.~J. Boersma.
\newblock Direct numerical simulations of turbulent flow over a permeable wall
  using a direct and a continuum approach.
\newblock {\em Phys. Fluids}, 17, 2005.

\bibitem{breugem:derivation:2005}
W.~P. Breugem and D.~A.~S. Rees.
\newblock A derivation of the volume--averaged {Boussinesq} equations for flow
  in porous media with viscous dissipation.
\newblock {\em Transp. Porous Media}, to appear.

\bibitem{canuto:spectral:1988}
C.~Canuto, M.~Hussaini, A.~Quarteroni, and T.~Zang.
\newblock {\em Spectral Methods in Fluid Dynamics}.
\newblock Springer Verlag (Berlin and New York), 1988.

\bibitem{carati:representation:1995}
D.~Carati, S.~Ghosal, and P.~Moin.
\newblock On the representation of backscatter in dynamic localization models.
\newblock {\em Phys. Fluids}, 7(3):606--616, 1995.

\bibitem{chasnov:simulation:1991}
J.~R. Chasnov.
\newblock Simulation of the {Kolmogorov} inertial subrange using an improved
  subgrid model.
\newblock {\em Phys. Fluids}, 3(1):188--200, 1991.

\bibitem{chen:high:1992}
S.~Chen and X.~Shan.
\newblock High-resolution turbulent simulations using the {Connection
  Machine-2}.
\newblock {\em Comput. Phys.}, 6(6):643--646, 1992.

\bibitem{cheng:near:2002}
H.~Cheng and I.~P. Castro.
\newblock Near wall flow over urban-like roughness.
\newblock {\em Boundary-Layer Meteorology}, 104(2):229--259, 2002.

\bibitem{eswaran:examination:1988}
V.~Eswaran and S.~B. Pope.
\newblock An examination of forcing in direct numerical simulations of
  turbulence.
\newblock {\em Comput. Fluids}, 16:257, 1988.

\bibitem{finnigan:turbulence:2000}
J.~Finnigan.
\newblock Turbulence in plant canopies.
\newblock {\em Ann. Rev. Fluid Mech.}, 32:519--571, 2000.

\bibitem{FFTW98}
M.~Frigo and S.~G. Johnson.
\newblock {FFTW}: An adaptive software architecture for the {FFT}.
\newblock In {\em Proc. 1998 IEEE Int. Conf. Acoustics Speech and Signal
  Processing}, volume~3, pages 1381--1384, http://www.fftw.org, 1998.

\bibitem{geurts:mixing:2001}
B.~J. Geurts.
\newblock Mixing efficiency in turbulent shear layers.
\newblock {\em J. Turbul.}, 2:2--24, 2001.

\bibitem{geurts:modern:2001}
B.~J. Geurts.
\newblock {\em Modern Simulation Strategies for Turbulent Flow}.
\newblock R.T. Edwards, 2001.

\bibitem{ghosal:dynamic:1995}
S.~Ghosal, T.~S. Lund, P.~Moin, and K.~Akselvoll.
\newblock A dynamic localization model for large-eddy simulation of turbulent
  flows.
\newblock {\em J. Fluid Mech.}, 286:229--255, 1995.

\bibitem{scsl}
Silicon Graphics.
\newblock {SCSL} user's guide.
\newblock TechPubs Library of Silicon Graphics, http://techpubs.sgi.com.

\bibitem{grossman:scaling:1992}
S.~Grossman and D.~Lohse.
\newblock Scaling in hard turbulent {Rayleight-Be\'nard} flow.
\newblock {\em Phys. Rev. A}, 46(2):903--917, 1992.

\bibitem{HDF5}
HDF5 Group.
\newblock {HDF5} user's guide.
\newblock National Center for Supercomputing Applications (NCSA), University of
  Illinois, http://www.hdfgroup.org.

\bibitem{hinze:turbulence:1975}
J.~O. Hinze.
\newblock {\em Turbulence: An Introduction to its Mechanism and Theory}.
\newblock New York, McGraw-Hill, 1975.

\bibitem{ishihara:high:2003}
T.~Ishihara and Y.~Kaneda.
\newblock High resolution {DNS} of incompressible homogeneous forced turbulence
  - time dependence of statistics.
\newblock In Y.~Kaneda and T.~Gotoh, editors, {\em Statistical Theories and
  Computational Approaches to Turbulence}. Springer, 2003.

\bibitem{jimenez:turbulent:2004}
J.~Jimenez.
\newblock Turbulent flow over rough walls.
\newblock {\em Ann. Rev. Fluid Mech.}, 36:173--196, 2004.

\bibitem{jimenez:structure:1993}
J.~Jimenez, A.~A. Wray, P.~G. Saffman, and R.~S. Rogallo.
\newblock The structure of intense vorticity in isotropic turbulence.
\newblock {\em J. Fluid Mech.}, 255:65--90, 1993.

\bibitem{kaneda:energy:2003}
Y.~Kaneda, T.~Ishihara, M.~Yokokawa, K.~Itakura, and A.~Uno.
\newblock Energy dissipation rate and energy spectrum in high resolution direct
  numerical simulations of turbulence in a periodic box.
\newblock {\em Phys. Fluids}, 15(2):21--24, 2003.

\bibitem{kerr:higher:1985}
R.~M. Kerr.
\newblock Higher-order derivative correlations and the alignment of small-scale
  structures in isotropic numerical turbulence.
\newblock {\em J. Fluid Mech.}, 153:31--58, 1985.

\bibitem{kerr:velocity:1990}
R.~M. Kerr.
\newblock Velocity, scalar and transfer spectra in numerical turbulence.
\newblock {\em J. Fluid Mech.}, 211:309--332, 1990.

\bibitem{kolmogorov:1941}
A.~N. Kolmogorov.
\newblock The local structure of turbulence in incompressible viscous fluids at
  very large {Reynolds} numbers.
\newblock {\em C.R. Acad. Sci. URSS}, 30:301--305, 1941.

\bibitem{kolmogorov:refinment:1962}
A.~N. Kolmogorov.
\newblock A refinement of previous hypothesis concerning the local structure of
  turbulence in a viscous incompressible fluid at high reynolds number.
\newblock {\em J. Fluid Mech.}, 13:82--85, 1962.

\bibitem{li:study:2005}
F.~Li, L.~Lefferts, and T.~H. van~der Meer.
\newblock Study on heat transfer enhancement by metallic foams with carbon nano
  fibers ({CNFs}).
\newblock In {\em Proceedings of the 6th World Conference on Experimental Heat
  Transfer, Fluid Mechanics, and Thermodynamics}, Matsushima, Miyagi, Japan,
  April 17-21, 2005.

\bibitem{machiels:predictability:1997}
L.~Machiels.
\newblock Predictability of small-scale motion in isotropic fluid turbulence.
\newblock {\em Phys. Rev. Lett.}, 79(18):3411--3414, 1997.

\bibitem{mazja:1985}
V.G. Maz'ja.
\newblock {\em Sobolev spaces}.
\newblock Springer Verlag -- Berling, 1985.

\bibitem{mazzi:fractal:2004}
B.~Mazzi and J.~C. Vassilicos.
\newblock Fractal generated turbulence.
\newblock {\em J. Fluid Mech.}, 502:65--87, 2004.

\bibitem{mccomb:physics:1990}
W.~D. McComb.
\newblock {\em The Physics of Fluid Turbulence}.
\newblock Oxford University Press, 1990.

\bibitem{meyers:accuracy:2004}
J.~Meyers.
\newblock {\em Accuracy of {Large-Eddy Simulation} strategies}.
\newblock PhD thesis, Katholieke Universiteit Leuven, 2004.

\bibitem{meyers:database:2003}
J.~Meyers, B.~J. Geurts, and M.~Baelmans.
\newblock Database analysis of errors in large-eddy simulation.
\newblock {\em Phys. Fluids}, 15(9):2740--2755, 2003.

\bibitem{misra:vortex:1997}
A.~Misra and D.~I. Pullin.
\newblock A vortex-based subgrid stress model for large-eddy simulation.
\newblock {\em Phys. Fluids}, 9(7):2443--2454, 1997.

\bibitem{mohseni:numerical:2003}
K.~Mohseni, B.~Kosovic, S.~Shkoller, and J.~E. Marsden.
\newblock Numerical simulations of the {Lagrangian Averaged Navier-Stokes}
  equations for homogeneous isotropic turbulence.
\newblock {\em Phys. Fluids}, 15(2):524--544, 2003.

\bibitem{MPI}
MPI.
\newblock {Message Passing Interface} standard.
\newblock http://www.mpi-forum.org.

\bibitem{pope:turbulent:2000}
S.~B. Pope.
\newblock {\em Turbulent Flows}.
\newblock Cambridge University Press, 2000.

\bibitem{rogallo:illiac:1977}
R.~S. Rogallo.
\newblock An {ILLIAC} program for the numerical simulation of homogeneous
  incompressible turbulence.
\newblock Technical Report NASA-TM-73203, NASA, 1977.

\bibitem{rogallo:numerical:1981}
R.~S. Rogallo.
\newblock Numerical experiments in homogeneous turbulence.
\newblock Technical Report NASA-TM-81315, NASA, 1981.

\bibitem{SARA}
SARA.
\newblock {Computing and Networking Services}.
\newblock http://www.sara.nl.

\bibitem{siggia:intermittency:1978}
E.~D. Siggia and G.~S. Patterson.
\newblock Intermittency effects in a numerical simulation of stationary
  three-dimensional turbulence.
\newblock {\em J. Fluid Mech.}, 86:567, 1978.

\bibitem{townsend:structure:1976}
A.~A. Townsend.
\newblock {\em The Structure of Turbulent Shear Flows}.
\newblock Cambridge Univ. Press, 1976.

\bibitem{vincent:spatial:1991}
A.~Vincent and M.~Meneguzzi.
\newblock The spatial structure and statistical properties of homogeneous
  turbulence.
\newblock {\em J. Fluid Mech.}, 225:1--20, 1991.

\bibitem{wang:examination:1996}
L.~P. Wang, S.~Chen, J.~G. Brasseur, and J.~C. Wyngaard.
\newblock Examination of hypotheses in the {Kolmogorov} refined turbulence
  theory through high-resolution simulations. {Part 1}. {Velocity field}.
\newblock {\em J. Fluid Mech.}, 309:113--156, 1996.

\bibitem{watanabe:statistics:2004}
T.~Watanabe and T.~Gotoh.
\newblock Statistics of a passive scalar in homogeneous turbulence.
\newblock {\em New J. Phys.}, 6(40):1--36, 2004.

\bibitem{wesseling:introduction:1992}
P.~Wesseling.
\newblock {\em An introduction to multigrid methods}.
\newblock Wiley, New York, 1992.

\bibitem{whitaker:forchkeimer:1996}
S.~Whitaker.
\newblock The {Forchheimer} equation: a theoretical development.
\newblock {\em Transp. Porous Media}, 25:27--61, 1996.

\bibitem{yamazaki:effects:2002}
Y.~Yamazaki, T.~Ishihara, and Y.~Kaneda.
\newblock Effects of wavenumber truncation on high-resolution direct numerical
  simulation of turbulence.
\newblock {\em J. Phys. Soc. Jpn.}, 71(3):777--781, 2002.

\bibitem{yeung:response:1991}
P.~K. Yeung and J.~G. Brasseur.
\newblock The response of isotropic turbulence to isotropic and anisotropic
  forcing at the large scales.
\newblock {\em Phys. Fluids A}, 3(5):884--897, 1991.

\bibitem{young:investigation:1999}
A.~J. Young.
\newblock {\em Investigation of Renormalization Group Methods for the Numerical
  Simulation of Isotropic Turbulence}.
\newblock Phd thesis, University of Edinburgh, 1999.

\end{thebibliography}

\end{document}